\documentclass[a4paper,11pt]{article}
\usepackage{jcappub} 

\arxivnumber{} 
\title{\boldmath Magnetic-field-induced chaos of charged particles around a rotating black hole in a King-type dark matter halo}

\author[a,b]{YuWang}
\author[b]{Meilin Liu}
\author[a]{Haiguang Xu}

\affiliation[a]{School of Physics and Astronomy, Shanghai Jiao Tong University, Shanghai 200240, China}

\affiliation[b]{School of Aeronautics and Astronautics, Shanghai Jiao Tong University, Shanghai 200240, China}

\emailAdd{sjtu2686361@sjtu.edu.cn}
\emailAdd{meilin.liu@sjtu.edu.cn}
\emailAdd{hgxu@sjtu.edu.cn}
\date{\today}

\abstract{We investigate the dynamics of charged test particles around
a rotating black hole embedded in a King-type dark matter
halo and subjected to a prescribed, asymptotically uniform
magnetic field. We adopt an effective rotating geometry
constructed from a static halo seed through a modified
Newman--Janis prescription and a test electromagnetic field
that generally requires supporting currents in the nonvacuum
background. In the absence of magnetic coupling, the timelike
Hamilton--Jacobi equation separates and admits a Carter-type
constant, providing an integrable reference system.
We characterize the orbital dynamics using Poincar\'e
sections and tangent-vector growth diagnostics, including
fast Lyapunov indicators and finite-time Lyapunov exponents,
with mass-shell monitoring and numerical refinement checks.
For the sampled initial conditions, increasing the magnetic
coupling can produce a transition from regular motion to
scattered Poincar\'e sections with sustained positive
finite-time growth rates. At fixed magnetic coupling, the two higher-density halo
cases yield larger finite-time exponents for a representative
chaotic orbit family, whereas the weak-halo correction is
not resolved consistently under numerical refinement. A different initial condition yields a closed Poincar\'e curve and a decreasing finite-time
exponent over an extended integration, supporting regular
motion despite the nonzero magnetic coupling. These results
show that the King-type halo modulates magnetic-field-induced
orbital instability in an initial-condition-dependent manner,
while regular motion persists within the magnetized system.
}

\begin{document}
\maketitle
\flushbottom

\section{Introduction}
\label{sec:introduction}

The motion of particles around black holes provides a direct
way to investigate the interplay between spacetime geometry
and dynamical stability. The Kerr solution describes the
gravitational field of an isolated rotating black hole
\cite{Kerr1963}, and its geodesic dynamics possesses a hidden
symmetry associated with the Carter constant
\cite{Carter1968}. Together with the conserved energy, axial
angular momentum, and mass-shell constraint, this additional
integral makes geodesic motion integrable. Environmental
fields can alter this structure, but a departure from the
vacuum Kerr geometry does not by itself imply
nonintegrability. Establishing the integrable limit of a
model is therefore essential before attributing irregular
trajectories to its surrounding matter distribution.

Dark matter is one component of the environments in which
astrophysical black holes reside. Halo models span a range
of central structures, including the cusped
Navarro--Frenk--White profile \cite{Navarro1997} and cored
prescriptions inspired by stellar-system models such as that
of King \cite{King1962}. The growth of a central black hole
can substantially reshape an initially specified
distribution: adiabatic growth may generate a density spike
\cite{Gondolo1999}, while a relativistic treatment modifies
the distribution in the strong-field region
\cite{Sadeghian2013}. Consequently, extrapolating a galactic
density law directly to the horizon requires additional
assumptions. In the present work, the term King-type refers
to a phenomenological cored density prescription with
specified inner and outer transitions, rather than a
self-consistent derivation of the original King distribution
function for dark matter near a rotating black hole.

The influence of dark matter on black-hole dynamics has
been explored through several complementary approaches.
Dark-matter minispikes can modify the gravitational-wave
evolution of extreme-mass-ratio systems \cite{Eda2013},
and relativistic black-hole solutions embedded in matter
distributions provide a framework for investigating
environmental effects on wave generation and propagation
\cite{Cardoso2022}. Effective black-hole geometries
associated with halo density profiles have also been
constructed \cite{Xu2018}. More recently, scalar
quasinormal modes have been studied in an effective rotating
background with a King-type cored halo \cite{Wang2026King}.
These investigations motivate an examination of nonlinear
particle dynamics as a complementary question: how does a
cored halo modify the phase-space structure of a system
perturbed by an external interaction?

To investigate this question, a rotating background must
be specified. The Newman--Janis algorithm and its
modifications provide useful prescriptions for constructing
rotating geometries from static seeds
\cite{NewmanJanis1965,Azreg2014}. However, the resulting
metric does not automatically inherit a consistent matter
interpretation, particularly outside the original setting
of the construction \cite{Hansen2013}. We therefore adopt
an effective rotating geometry derived from a static
King-type halo seed and examine its Hamilton--Jacobi
separability explicitly. Its interpretation and the
assumptions concerning the matter source are discussed
in Sec.~2.

Magnetic fields supply a physically motivated perturbation
of charged-particle motion. The Wald solution describes an
asymptotically uniform, aligned electromagnetic test field
around a stationary, axisymmetric vacuum black hole
\cite{Wald1974}, while the Blandford--Znajek mechanism
demonstrates the broader importance of electromagnetic
fields in extracting rotational energy from black holes
\cite{Blandford1977}. An asymptotically uniform-field model
is an idealization of a black-hole environment, rather
than a complete description of an accretion-flow
magnetosphere. It nevertheless permits a controlled study
of the competition between gravitational and Lorentz
forces. In the test-field approximation, the
electromagnetic stress-energy is neglected in the geometry,
although the charge-to-mass ratio can make the Lorentz
force dynamically important for the particle.

Chaotic charged-particle dynamics in black-hole
electromagnetic fields has an established literature.
Takahashi and Koyama investigated spin-dependent chaotic
motion in a dipolar magnetic field around a rotating
black hole \cite{Takahashi2009}. Transitions between regular
and chaotic motion in magnetized coronae were studied using
phase-space and recurrence diagnostics \cite{Kopacek2010},
and the influence of breaking axial symmetry was examined
for inclined magnetic fields \cite{Kopacek2014}.
Chaotic scattering can also transfer energy between
different components of particle motion and contribute
to acceleration along magnetic-field lines
\cite{Stuchlik2016}. Explicit symplectic integration has
subsequently been applied to charged-particle chaos around
a magnetized Kerr black hole \cite{Sun2021}. These results
establish that magnetic-field-induced chaos does not
require a dark-matter halo.

Recent studies have extended this problem to nonvacuum
and modified black-hole backgrounds. Cao et al. investigated
electromagnetic fields and chaotic charged-particle motion
around hairy black holes in Horndeski gravity
\cite{Cao2024}. Lu and Wu studied magnetic-field-induced
chaos near a renormalization-group-improved Kerr black
hole \cite{Lu2025}, while Xu et al. analyzed charged-particle
dynamics in magnetized Kerr-MOG geometries
\cite{Xu2025MOG,Xu2026Potential}. The electromagnetic
prescription is particularly important in such extensions,
since a Wald-type potential constructed from Killing
vectors is not generally a source-free Maxwell solution
in a nonvacuum geometry. Here it is treated as a prescribed
external test field with potentially nonzero supporting
currents. Charged-particle oscillations in black-hole
backgrounds with cold-dark-matter halos have also been
investigated in connection with high-frequency
quasi-periodic oscillations \cite{Ahal2025}.
These developments motivate a more specific question:
at fixed asymptotic magnetic coupling, how does a
King-type cored halo affect orbital instability, and
under what initial conditions does regular motion persist?

Reliable characterization of this dependence requires
both geometrical and dynamical diagnostics. Poincar\'e
sections reveal the organization of trajectories into
invariant curves, resonance structures, and irregular
regions, following the classical use of surfaces of
section in nonlinear Hamiltonian dynamics
\cite{Henon1964}. Lyapunov methods quantify the growth
of perturbations \cite{Benettin1976}, and fast Lyapunov
indicators provide a practical way to distinguish different
dynamical regimes \cite{Froeschle1997}. Their implementation
in relativistic systems requires a specified time parameter
and a consistent definition of separation \cite{Wu2006}.
Moreover, finite-time estimates must be distinguished
from asymptotic exponents, particularly near weakly chaotic
or sticky trajectories \cite{Skokos2010}. We therefore
combine Poincar\'e sections with proper-time tangent-vector
growth diagnostics, supplemented by mass-shell monitoring
and numerical refinement checks. The phase-space norm
and the relation between the fast Lyapunov indicator
and the finite-time exponent are specified in Sec.~3.

In this work, we establish the separable,
zero-magnetic-coupling limit of the adopted rotating
King-type background and use its Carter-type integral
as a reference for the numerical calculations.
We then examine regular and chaotic charged-particle
motion as the magnetic coupling is varied.
At fixed coupling, comparisons with the vacuum
background and scans over halo density and initial
radius reveal an orbit-dependent response:
higher densities increase the finite-time growth
rates of a representative chaotic family, while
extended integrations of another initial condition
support persistent regular motion over the computed
interval. These comparisons are performed at fixed
canonical energy and angular momentum, with the initial
polar momentum determined by the mass-shell constraint.
The resulting analysis characterizes both the enhancement
of orbital instability and the persistence of regular
structures within the adopted magnetized halo model.

The paper is organized as follows. Section~2 presents
the static halo seed, the effective rotating geometry,
and the electromagnetic prescription. Section~3 introduces
the Hamiltonian equations, the integrable limit, and
the numerical diagnostics. Section~4 examines
magnetic-field-induced chaos. Section~5 investigates
the influence of the halo parameters and initial
conditions. Section~6 summarizes the results and
discusses the limitations of the model.
Throughout this work, we use geometrized units
$G=c=1$ and normalize lengths and times by the
central mass parameter $M$ of the static seed,
unless otherwise stated.

\section{Background geometry and magnetic field}
\label{sec:background}

\subsection{King-type density profile and static seed}
\label{subsec:static_seed}

We adopt the cored density prescription
\begin{equation}
\rho(r)=
\frac{\rho_s}{[1+(r/r_s)^2]^{3/2}}\,W(r),
\label{eq:king_density}
\end{equation}
where $\rho_s$ sets the density normalization and
$r_s$ is the core scale.
This phenomenological King-type prescription is
inspired by cored stellar-system models
\cite{King1962}; it is not identified with an exact
lowered-isothermal distribution function in the
black-hole spacetime.
The window function $W(r)$ removes matter from
the inner region and gives the halo a finite
outer extent. These transitions are prescribed
parts of the model rather than consequences of
the collisionless evolution of dark matter.

To match the numerical implementation, we define
\begin{equation}
S(u)=
\begin{cases}
0, & u\leq0,\\
6u^5-15u^4+10u^3, & 0<u<1,\\
1, & u\geq1,
\end{cases}
\label{eq:smooth_step}
\end{equation}
and
\begin{equation}
\begin{split}
W(r)={}&
S\!\left(
\frac{r-r_{\rm in}}{\delta_{\rm in}}
\right)\\
&\times
\left[
1-S\!\left(
\frac{r-r_{\rm out}}{\delta_{\rm out}}
\right)
\right].
\end{split}
\label{eq:halo_window}
\end{equation}
The window function and its first two derivatives
are continuous across the transition boundaries.
The calculations use
\begin{equation}
\begin{split}
r_{\rm in}&=8M,\qquad
\delta_{\rm in}=2M,\\
r_{\rm out}&=200M,\qquad
\delta_{\rm out}=20M.
\end{split}
\label{eq:cutoff_parameters}
\end{equation}
Thus, the density vanishes for $r\leq8M$,
follows the unmodified King-type law over
$10M\leq r\leq200M$, and vanishes again for
$r\geq R_{\rm h}=220M$.
Accordingly, $\rho_s$ is a density normalization
rather than the actual central density of the
truncated configuration.
In geometrized units, $\rho_s$ has dimensions
of $M^{-2}$, and the corresponding dimensionless
parameter is $\rho_sM^2$.

The static line element is
\begin{equation}
\begin{split}
ds_0^2={}&-A(r)\,dt^2
+\frac{dr^2}{B(r)}\\
&+r^2\left(
d\theta^2+\sin^2\theta\,d\phi^2
\right),
\end{split}
\label{eq:static_metric}
\end{equation}
with signature $(-,+,+,+)$ and
\begin{equation}
B(r)=1-\frac{2m(r)}{r}.
\label{eq:static_B}
\end{equation}
We specify the static stress tensor by
\begin{equation}
T^\mu{}_\nu
=\operatorname{diag}
[-\rho(r),0,p_\perp(r),p_\perp(r)],
\label{eq:static_stress}
\end{equation}
so that the radial pressure vanishes.
This is an anisotropic-fluid closure of the
static problem, of the type relevant to
relativistic matter-supported black-hole
models \cite{Cardoso2022}.
It does not, by itself, specify a microscopic
dark-matter distribution function.

The independent Einstein equations give
\begin{equation}
m'(r)=4\pi r^2\rho(r),
\label{eq:mass_equation}
\end{equation}
and
\begin{equation}
\frac{A'(r)}{A(r)}
=\frac{2m(r)}{r[r-2m(r)]},
\label{eq:lapse_equation}
\end{equation}
where a prime denotes differentiation with
respect to $r$.
Taking $m(r_{\rm in})=M$, we obtain
\begin{equation}
m(r)=M+4\pi\int_{r_{\rm in}}^r
x^2\rho(x)\,dx
\label{eq:mass_integral}
\end{equation}
for $r\geq r_{\rm in}$, while $m(r)=M$
in the inner vacuum region.
Here $M$ is the central mass parameter of
the static seed.
The total mass and halo contribution are
\begin{equation}
M_{\rm tot}=m(R_{\rm h}),
\qquad
M_{\rm h}=M_{\rm tot}-M.
\label{eq:total_mass}
\end{equation}

Normalizing the time coordinate at infinity
fixes
\begin{equation}
A(R_{\rm h})=B(R_{\rm h})
=1-\frac{2M_{\rm tot}}{R_{\rm h}},
\label{eq:lapse_boundary}
\end{equation}
and hence, for $r_{\rm in}\leq r\leq R_{\rm h}$,
\begin{equation}
\begin{split}
A(r)={}&
\left(
1-\frac{2M_{\rm tot}}{R_{\rm h}}
\right)\\
&\times
\exp\!\left[
-\int_r^{R_{\rm h}}
\frac{2m(x)}{x[x-2m(x)]}\,dx
\right].
\end{split}
\label{eq:lapse_integral}
\end{equation}
Outside the halo,
$A=B=1-2M_{\rm tot}/r$.
Within the matter distribution, $A$ and $B$
are generally different; their equality is
not imposed.

Stress-energy conservation determines the
tangential pressure:
\begin{equation}
\begin{split}
p_\perp(r)
&=\frac{\rho(r)\,r}{4}
\frac{A'(r)}{A(r)}\\
&=\frac{\rho(r)m(r)}
{2[r-2m(r)]}.
\end{split}
\label{eq:tangential_pressure}
\end{equation}
The static construction is restricted to
$r>2m(r)$ outside the central horizon.
Parameter choices that introduce an additional
zero of $B(r)$ in the halo require a separate
treatment and are excluded from this setup.

The smooth transitions in
Eq.~\eqref{eq:halo_window} avoid discontinuous
matter boundaries.
In the inner vacuum region, $2M<r\leq r_{\rm in}$,
the lapse takes the form
\begin{equation}
A(r)=C\left(1-\frac{2M}{r}\right),
\label{eq:inner_lapse}
\end{equation}
where
\begin{equation}
C=
\frac{A(r_{\rm in})}
{1-2M/r_{\rm in}}.
\label{eq:inner_redshift}
\end{equation}
This constant redshift factor is determined
by the exterior halo and is generally
different from unity.

\subsection{Effective rotating geometry}
\label{subsec:rotating_geometry}

Following the modified Newman--Janis approach
\cite{NewmanJanis1965,Azreg2014}, we introduce
\begin{equation}
\begin{split}
K(r)&=r^2\sqrt{\frac{B(r)}{A(r)}},\\
\Delta(r)&=r^2B(r)+a^2,\\
\mathcal{S}(r,\theta)
&=K(r)+a^2\cos^2\theta,\\
\Psi(r,\theta)
&=r^2+a^2\cos^2\theta,
\end{split}
\label{eq:rotating_functions}
\end{equation}
where $a$ has dimensions of length.
The effective line element is
\begin{equation}
\begin{split}
ds^2={}&
-\frac{\Psi\Delta}{\mathcal{S}^2}
\left(dt-a\sin^2\theta\,d\phi\right)^2\\
&+\frac{\Psi}{\Delta}\,dr^2
+\Psi\,d\theta^2\\
&+\frac{\Psi\sin^2\theta}{\mathcal{S}^2}
\left[a\,dt-(K+a^2)d\phi\right]^2.
\end{split}
\label{eq:rotating_metric}
\end{equation}
The functions $\Psi$ and $\mathcal{S}$ are
distinct whenever $A\neq B$.
In particular, replacing $K$ by $r^2$
inside the halo would change the model.

The choice of $\Psi$ in
Eq.~\eqref{eq:rotating_functions} is part
of the effective rotating prescription.
It recovers the static seed and permits
separation of the massive Hamilton--Jacobi
equation in the absence of electromagnetic
coupling, as shown in Sec.~3.
It is not determined solely by the density
profile.

A Newman--Janis construction does not
automatically preserve a specified matter
model \cite{Hansen2013}.
We therefore do not assume that the rotating
stress tensor retains the static form in
Eq.~\eqref{eq:static_stress}.
The metric is used as an effective rotating
background, without claiming an independently
validated rotating Einstein--dark-matter
solution with a specified distribution
function.

Two limiting cases provide useful checks.
Setting $a=0$ in
Eq.~\eqref{eq:rotating_metric} gives
\begin{equation}
\begin{split}
ds^2\big|_{a=0}
&=-\frac{r^4B}{K^2}\,dt^2
+\frac{dr^2}{B}+r^2d\Omega^2\\
&=-A\,dt^2+\frac{dr^2}{B}+r^2d\Omega^2,
\end{split}
\label{eq:static_limit}
\end{equation}
which recovers Eq.~\eqref{eq:static_metric}.
Setting $\rho_s=0$ instead yields
\begin{equation}
\begin{split}
A=B&=1-\frac{2M}{r},
\qquad K=r^2,\\
\Delta&=r^2-2Mr+a^2,
\qquad \mathcal{S}=\Psi,
\end{split}
\label{eq:kerr_limit}
\end{equation}
and therefore recovers the Kerr metric.

For $r\geq R_{\rm h}$, the geometry has
the Kerr form with mass $M_{\rm tot}$
and angular momentum $J=aM_{\rm tot}$.
The parameter $a/M$ used in the numerical
calculations is normalized by the seed mass;
it should not be confused with the
dimensionless asymptotic spin
$J/M_{\rm tot}^2=a/M_{\rm tot}$.

For later use, the nontrivial inverse
metric components are
\begin{equation}
\begin{split}
g^{tt}
&=-\frac{
(K+a^2)^2-a^2\Delta\sin^2\theta
}{\Psi\Delta},\\
g^{t\phi}
&=-\frac{a(K+a^2-\Delta)}
{\Psi\Delta},\\
g^{\phi\phi}
&=\frac{\Delta-a^2\sin^2\theta}
{\Psi\Delta\sin^2\theta},\\
g^{rr}
&=\frac{\Delta}{\Psi},
\qquad
g^{\theta\theta}=\frac{1}{\Psi}.
\end{split}
\label{eq:inverse_metric}
\end{equation}
These expressions are evaluated on the
regular exterior coordinate domain used
for the orbital calculations.

\subsection{Prescribed magnetic test field}
\label{subsec:magnetic_field}

Let $\xi^\mu=(\partial_t)^\mu$ and
$\psi^\mu=(\partial_\phi)^\mu$ denote
the stationary and axial Killing vectors.
We prescribe a Wald-type potential
\cite{Wald1974},
\begin{equation}
A_\mu
=\frac{B_0}{2}
\left(g_{\mu\phi}+2a g_{\mu t}\right)
+aB_0\,\delta^t_\mu,
\label{eq:wald_potential}
\end{equation}
where $B_0$ controls the asymptotic
magnetic-field strength and orientation.
No additional black-hole electric-charge
term is included.
In the vacuum Kerr limit, this corresponds
to the uncharged-hole choice rather than
the equilibrium Wald-charge configuration.

The only nonzero components are
\begin{equation}
\begin{split}
A_t
&=\frac{B_0}{2}
\left(g_{t\phi}+2a g_{tt}\right)+aB_0,\\
A_\phi
&=\frac{B_0}{2}
\left(g_{\phi\phi}+2a g_{t\phi}\right).
\end{split}
\label{eq:potential_components}
\end{equation}
The additive constant fixes $A_t\to0$
at spatial infinity without changing
the field tensor,
\begin{equation}
F_{\mu\nu}
=\partial_\mu A_\nu-\partial_\nu A_\mu.
\label{eq:field_tensor}
\end{equation}
In the asymptotic region,
\begin{equation}
A_\phi\sim
\frac{B_0}{2}r^2\sin^2\theta,
\label{eq:asymptotic_potential}
\end{equation}
corresponding to a uniform magnetic field
aligned with the rotation axis.
For $a\neq0$, the potential also contains
a nontrivial electric component.

The Killing-vector construction yields
a source-free Maxwell test field in a
Ricci-flat background.
This property does not generally extend
to a halo-supported geometry, as emphasized
in studies of electromagnetic fields around
nonvacuum black holes
\cite{Cao2024,Xu2026Potential}.
We therefore allow a supporting current.

With the metric signature $(-,+,+,+)$,
the field-tensor definition in
Eq.~\eqref{eq:field_tensor}, and the particle
coupling convention $p_\mu-qA_\mu$,
we use Gaussian electromagnetic conventions
and write
\begin{equation}
J^\mu
=\frac{1}{4\pi}\nabla_\nu F^{\nu\mu}.
\label{eq:supporting_current}
\end{equation}
No condition $J^\mu=0$ is imposed
throughout the halo-supported background.
The prescribed field represents an external
electromagnetic environment; its supporting
current is not identified with a dark-matter
charge current.

The electromagnetic stress-energy is
neglected in constructing the metric.
This approximation requires the field energy
to be sufficiently small compared with the
background curvature scale over the orbital
domain.
An exactly uniform magnetic field extending
to infinity would not admit globally
negligible gravitational backreaction.
Its asymptotic form here is therefore
understood as part of a test-field
idealization applicable to a finite
physical region.

\subsection{Magnetic coupling and comparison conventions}
\label{subsec:coupling}

For a test particle with rest mass $m_{\rm p}$
and charge $q$, we define
\begin{equation}
\epsilon=\frac{q}{m_{\rm p}},
\qquad
\beta=\epsilon B_0M.
\label{eq:magnetic_coupling}
\end{equation}
There is no factor of $1/2$ in the
definition of $\beta$.
The charge-weighted potential is
$\mathcal{A}_\mu=\epsilon A_\mu$.
Before setting $M=1$, it reads
\begin{equation}
\begin{split}
\mathcal{A}_\mu
={}&\frac{\beta}{2M}
\left(g_{\mu\phi}+2a g_{\mu t}\right)\\
&+\frac{a\beta}{M}\delta^t_\mu.
\end{split}
\label{eq:effective_potential_dimensional}
\end{equation}
In the numerical units $M=1$, its
nonzero components reduce to
\begin{equation}
\begin{split}
\mathcal{A}_t
&=\frac{\beta}{2}
\left(g_{t\phi}+2a g_{tt}\right)+a\beta,\\
\mathcal{A}_\phi
&=\frac{\beta}{2}
\left(g_{\phi\phi}+2a g_{t\phi}\right).
\end{split}
\label{eq:effective_potential}
\end{equation}

For a fixed coordinate orientation,
the sign of $\beta$ specifies the sign
of the charge--field product.
Negligible gravitational backreaction
of the field does not require a negligible
Lorentz force, because the latter also
depends on the particle charge-to-mass ratio.
Accordingly, $\beta$ alone does not determine
whether the test-field approximation is
valid: the field strength and the
charge-to-mass ratio must be considered
separately.

All comparisons use the gauge specified
in Eq.~\eqref{eq:potential_components}.
Since the potential depends on the metric,
fixing $\beta$ while varying the halo
density holds the asymptotic coupling fixed,
but does not hold the locally measured
electromagnetic field fixed everywhere.
The density scans therefore compare
gravitational and prescribed electromagnetic
backgrounds under a common asymptotic
field prescription.

Radiation reaction, self-force, dynamical
friction, and electromagnetic backreaction
are omitted.
The following sections concern conservative
charged-test-particle dynamics within this
approximation.

\section{Hamiltonian dynamics, integrable limit, and numerical diagnostics}
\label{sec:dynamics}

We describe the motion of a charged test particle
using a Hamiltonian formulation with proper time
$\tau$ as the evolution parameter.
Momenta and conserved quantities are normalized
by the particle rest mass.
The distinction between canonical and kinetic
momenta is retained throughout, since the
electromagnetic potential contributes to the
canonical energy and angular momentum.

\subsection{Hamiltonian and equations of motion}
\label{subsec:hamiltonian}

Let $p_\mu$ denote the specific canonical momentum.
The kinetic momentum is
\begin{equation}
\pi_\mu=p_\mu-\mathcal{A}_\mu,
\qquad
\mathcal{A}_\mu=\frac{q}{m_{\rm p}}A_\mu.
\label{eq:kinetic_momentum}
\end{equation}
For a massive particle, $\pi_\mu=u_\mu$,
where $u^\mu=dx^\mu/d\tau$.
The Hamiltonian is
\begin{equation}
H=\frac{1}{2}g^{\mu\nu}\pi_\mu\pi_\nu,
\qquad
H=-\frac{1}{2}
\label{eq:charged_hamiltonian}
\end{equation}
on the timelike mass shell.

Stationarity and axial symmetry of both the
metric and the chosen potential imply
\begin{equation}
p_t=-E,\qquad p_\phi=L,
\label{eq:canonical_constants}
\end{equation}
where $E$ and $L$ are the specific canonical
energy and axial angular momentum.
These constants refer to the gauge fixed
in Sec.~2.
For nonzero electromagnetic coupling,
they must not be identified with $-u_t$
and $u_\phi$.

It is convenient to introduce
\begin{equation}
\begin{split}
e(r,\theta)
&=E+\mathcal{A}_t(r,\theta),\\
\ell(r,\theta)
&=L-\mathcal{A}_\phi(r,\theta),\\
\mathcal{P}(r,\theta)
&=(K+a^2)e-a\ell,\\
\mathcal{W}(r,\theta)
&=\frac{\ell}{\sin\theta}
-ae\sin\theta.
\end{split}
\label{eq:auxiliary_momenta}
\end{equation}
Thus $\pi_t=-e$ and $\pi_\phi=\ell$,
while neither $e$ nor $\ell$ is generally
conserved.
Since $\mathcal{A}_r=\mathcal{A}_\theta=0$
in the adopted gauge, the radial and polar
canonical momenta coincide with their
kinetic counterparts.

Substituting Eq.~\eqref{eq:inverse_metric}
into Eq.~\eqref{eq:charged_hamiltonian}
gives the reduced Hamiltonian
\begin{equation}
H=\frac{1}{2\Psi}
\left[
\Delta p_r^2+p_\theta^2
-\frac{\mathcal{P}^2}{\Delta}
+\mathcal{W}^2
\right].
\label{eq:reduced_hamiltonian}
\end{equation}
In particular, the mixed contribution in
the inverse-metric representation is
$-2g^{t\phi}e\ell$, because $\pi_t=-e$.

Define the numerator
\begin{equation}
\mathcal{N}
=\Delta p_r^2+p_\theta^2
-\frac{\mathcal{P}^2}{\Delta}
+\mathcal{W}^2.
\label{eq:hamiltonian_numerator}
\end{equation}
For $y=(r,\theta,p_r,p_\theta)$,
Hamilton's equations become
\begin{equation}
\begin{split}
\dot r
&=\frac{\Delta p_r}{\Psi},\\
\dot\theta
&=\frac{p_\theta}{\Psi},\\
\dot p_r
&=-\frac{\mathcal{N}_{,r}}{2\Psi}
+\frac{r\mathcal{N}}{\Psi^2},\\
\dot p_\theta
&=-\frac{\mathcal{N}_{,\theta}}{2\Psi}
-\frac{a^2\sin\theta\cos\theta}{\Psi^2}
\mathcal{N},
\end{split}
\label{eq:reduced_equations}
\end{equation}
where a dot denotes differentiation with
respect to $\tau$.
The partial derivatives are evaluated at
fixed momenta and fixed $E,L$.
Explicitly,
\begin{equation}
\begin{split}
\mathcal{N}_{,r}={}&
\Delta' p_r^2
-\frac{2\mathcal{P}\mathcal{P}_{,r}}{\Delta}\\
&+\frac{\mathcal{P}^2\Delta'}{\Delta^2}
+2\mathcal{W}\mathcal{W}_{,r},\\
\mathcal{N}_{,\theta}={}&
-\frac{2\mathcal{P}\mathcal{P}_{,\theta}}{\Delta}
+2\mathcal{W}\mathcal{W}_{,\theta}.
\end{split}
\label{eq:numerator_derivatives}
\end{equation}
All derivatives of the electromagnetic
potential are included through $e$ and
$\ell$.
In particular,
$e_{,i}=\mathcal{A}_{t,i}$ and
$\ell_{,i}=-\mathcal{A}_{\phi,i}$
for $i=r,\theta$.

The Hamiltonian derivatives are evaluated
before imposing the mass-shell condition.
Although $\mathcal{N}=-\Psi$ on a physical
trajectory, this relation must not be
substituted into the Hamiltonian before
taking phase-space derivatives.

The cyclic coordinates can be reconstructed
from
\begin{equation}
\begin{split}
\dot t
&=\frac{1}{\Psi}
\left[
\frac{(K+a^2)\mathcal{P}}{\Delta}
+a\mathcal{W}\sin\theta
\right],\\
\dot\phi
&=\frac{1}{\Psi}
\left[
\frac{a\mathcal{P}}{\Delta}
+\frac{\mathcal{W}}{\sin\theta}
\right].
\end{split}
\label{eq:cyclic_equations}
\end{equation}
They do not enter the autonomous reduced
system.
We therefore integrate the four-dimensional
reduced phase space for the Poincar\'e
and tangent-vector analyses.

\subsection{Separable limit and Carter-type integral}
\label{subsec:integrable_limit}

When $\beta=0$, the charge-weighted
potential vanishes, so that $e=E$ and
$\ell=L$.
Let $H_0$ denote the Hamiltonian in this
limit.
Following the Hamilton--Jacobi separation
underlying the Carter integral
\cite{Carter1968}, we use the specific
principal function
\begin{equation}
S=-Et+L\phi+S_r(r)+S_\theta(\theta)
+\frac{\tau}{2}.
\label{eq:HJ_ansatz}
\end{equation}
The Hamilton--Jacobi equation,
$\partial_\tau S+H_0(x,\partial S)=0$,
then becomes
\begin{equation}
\begin{split}
0={}&
\Delta\left(\frac{dS_r}{dr}\right)^2
-\frac{[(K+a^2)E-aL]^2}{\Delta}
+r^2\\
&+\left(\frac{dS_\theta}{d\theta}\right)^2
+\left(\frac{L}{\sin\theta}
-aE\sin\theta\right)^2
+a^2\cos^2\theta.
\end{split}
\label{eq:HJ_separation}
\end{equation}
Since $K$ and $\Delta$ depend only on $r$,
the first line is radial and the second
line is angular.

A convenient separation constant is
\begin{equation}
Q=p_\theta^2+\cos^2\theta
\left[
a^2(1-E^2)+\frac{L^2}{\sin^2\theta}
\right].
\label{eq:carter_constant}
\end{equation}
The separated potentials are
\begin{equation}
\begin{split}
\mathcal{R}(r)={}&
[(K+a^2)E-aL]^2\\
&-\Delta
\left[r^2+(L-aE)^2+Q\right],\\
\Theta(\theta)={}&
Q-\cos^2\theta
\left[
a^2(1-E^2)+\frac{L^2}{\sin^2\theta}
\right],
\end{split}
\label{eq:separated_potentials}
\end{equation}
so that
\begin{equation}
\Psi^2\dot r^{\,2}=\mathcal{R}(r),
\qquad
\Psi^2\dot\theta^{\,2}=\Theta(\theta).
\label{eq:separated_motion}
\end{equation}

For an off-shell phase-space expression,
the unit rest-mass term in
Eq.~\eqref{eq:carter_constant} is replaced
by $-2H_0$:
\begin{equation}
Q_{\rm ext}
=p_\theta^2+\cos^2\theta
\left[
\frac{L^2}{\sin^2\theta}
-a^2E^2-2a^2H_0
\right].
\label{eq:extended_carter}
\end{equation}
It reduces to $Q$ on the timelike
mass shell.
The separation supplies the additional
integral needed for Liouville integrability,
with functional independence understood
at generic phase-space points.

This result holds for the adopted radial
functions $K(r)$ and $\Delta(r)$ irrespective
of the particular halo parameters, provided
the background remains regular on the
domain considered.
Thus, within this metric prescription,
increasing the King halo density does not
by itself make neutral geodesic motion
nonintegrable.
This statement depends on the chosen
rotating geometry and conformal factor;
it is not a general claim about all
rotating dark-matter configurations.

For $\beta\neq0$, the functions $e$ and
$\ell$ acquire coupled radial and angular
dependence, and the above separation is
generally lost.
We evaluate the neutral expression along
charged trajectories as
\begin{equation}
Q_{\rm ref}(\tau)
=p_\theta^2+\cos^2\theta
\left[
a^2(1-E^2)+\frac{L^2}{\sin^2\theta}
\right].
\label{eq:reference_carter}
\end{equation}
Its variation measures the failure of
this particular neutral integral to
remain conserved.
It is neither a numerical constraint error
nor a sufficient criterion for chaos:
regular trajectories may persist in the
magnetically coupled system even when
$Q_{\rm ref}$ varies.
The loss of the displayed separation alone
also does not establish the absence of
every possible additional integral.

\subsection{Initial conditions and Poincar\'e sections}
\label{subsec:initial_sections}

We choose equatorial initial conditions
with vanishing initial radial velocity,
\begin{equation}
\theta_0=\frac{\pi}{2},
\qquad
p_{r0}=0,
\label{eq:initial_position}
\end{equation}
and specify $E,L,r_0$ and the background
parameters.
The remaining momentum follows from
the timelike mass-shell condition:
\begin{equation}
p_{\theta0}
=
\left[
\frac{\mathcal{P}_0^2}{\Delta_0}
-\mathcal{W}_0^2-\Psi_0
-\Delta_0p_{r0}^2
\right]^{1/2}.
\label{eq:initial_ptheta}
\end{equation}
The positive branch selects motion toward
increasing $\theta$.
We retain strictly positive radicands for
the section studies, excluding degenerate
equatorial launches.

The mass-shell condition is necessary
but does not replace the requirement that
a physical trajectory be future-directed.
On an exterior chart where $t$ is a time
function, future-directed motion requires
$\dot t>0$, with $\dot t$ given by
Eq.~\eqref{eq:cyclic_equations}.

The main comparisons use $a/M=0.9$,
$r_s/M=10$, $E=0.97$, and $L/M=3.8$,
with $r_0$, $\rho_s$, and $\beta$ varied
as specified in Secs.~4 and 5.
At fixed $E,L,r_0$, changing the magnetic
or halo parameters also changes
$p_{\theta0}$ through
Eq.~\eqref{eq:initial_ptheta}.
The resulting scans therefore describe
families with fixed canonical constants
and specified coordinates, rather than
identical initial local velocities.

The Poincar\'e surface of section is
defined by
\begin{equation}
\theta=\frac{\pi}{2},
\qquad
\dot\theta>0,
\label{eq:poincare_surface}
\end{equation}
and the intersections are plotted in
the $(r,p_r)$ plane.
Crossings are located using event detection
and the numerical solution's continuous
interpolant; the initial crossing at
$\tau=0$ is omitted.

Smooth invariant curves are consistent
with quasiperiodic motion, while resolved
island chains indicate resonant structures.
Scattered intersections distributed over
a two-dimensional region can indicate
chaotic motion
\cite{Henon1964,Kopacek2010}.
However, a sparsely sampled or incomplete
curve is not sufficient, on its own,
to distinguish these possibilities.
Section morphology is therefore interpreted
together with the perturbation-growth
diagnostics.

\subsection{Tangent dynamics and finite-time indicators}
\label{subsec:lyapunov}

Writing Eq.~\eqref{eq:reduced_equations}
as $\dot y=f(y)$, the tangent vector $v$
obeys
\begin{equation}
\dot v=\mathbf{J}_f(y)v,
\qquad
(\mathbf{J}_f)_{ij}
=\frac{\partial f_i}{\partial y_j}.
\label{eq:variational_equations}
\end{equation}
The orbit and tangent equations are
integrated together with the same
proper-time parameter, avoiding comparisons
between independent adaptive time grids.

In the dimensionless numerical variables
with $M=1$, we use the fixed scale matrix
\begin{equation}
\mathbf{D}
=\operatorname{diag}(10,1,0.1,1)
\label{eq:tangent_scales}
\end{equation}
and the norm
\begin{equation}
\|v\|_D=\|\mathbf{D}^{-1}v\|_2.
\label{eq:tangent_norm}
\end{equation}
The initial deviation is chosen tangent
to the fixed-$E,L$ mass shell.
Defining
$g=\mathbf{D}^{T}\nabla_y H(y_0)$,
we project a trial direction $w_*$
in scaled coordinates according to
\begin{equation}
\begin{split}
w_\perp
&=w_*-g\frac{g^Tw_*}{g^Tg},\\
v_0
&=\mathbf{D}
\frac{w_\perp}{\|w_\perp\|_2},
\end{split}
\label{eq:tangent_projection}
\end{equation}
for a nonzero constraint gradient and a
nondegenerate projected direction.
The calculations use
\begin{equation}
w_*=(1,0.3,0.2,0.1)^T.
\label{eq:initial_trial_direction}
\end{equation}
Thus $\nabla H\cdot v_0=0$ and
$\|v_0\|_D=1$.
The tangent vector represents an
infinitesimal perturbation; its unit
normalization does not represent a
finite displacement of the orbit.

Following standard renormalized Lyapunov
calculations \cite{Benettin1976,Skokos2010},
we periodically compute
\begin{equation}
s_k=\|v(\tau_k^-)\|_D,
\qquad
v(\tau_k^+)=\frac{v(\tau_k^-)}{s_k}.
\label{eq:renormalization}
\end{equation}
The accumulated logarithmic growth is
\begin{equation}
\Lambda(\tau_n)
=\sum_{k=1}^{n}\ln s_k.
\label{eq:accumulated_growth}
\end{equation}
At the final time $T$, including any
shortened final renormalization interval,
we report
\begin{equation}
\lambda_T=\frac{\Lambda(T)}{T},
\qquad
\mathrm{FLI}(T)=\frac{\Lambda(T)}{\ln10}.
\label{eq:finite_time_indicators}
\end{equation}

Here $\mathrm{FLI}$ denotes a renormalized
logarithmic fast Lyapunov indicator
\cite{Froeschle1997}.
This definition follows perturbation
growth without overflow and is not a
running-maximum variant of the indicator.
The two quantities satisfy
$\mathrm{FLI}=T\lambda_T/\ln10$
and therefore are not independent
diagnostics.

The finite-time estimate is obtained
from one evolved tangent direction.
Its interpretation as an estimate of
the largest asymptotic Lyapunov exponent
relies on the alignment of a generic
initial direction with the dominant
growing direction.
At finite times, the result can retain
a dependence on the initial direction.

Sustained exponential growth is associated
with an exponent approaching a positive
plateau, whereas algebraic growth along
a regular trajectory gives an exponent
that decreases toward zero.
These asymptotic behaviors can be
difficult to distinguish over a short
interval, particularly for sticky
trajectories.
We consequently do not adopt a universal
numerical FLI threshold.
The time dependence of the indicators
is interpreted together with the
Poincar\'e sections and numerical
sensitivity checks.

Proper time and the phase-space norm
are explicitly specified because
relativistic Lyapunov diagnostics
require care in their definition
\cite{Wu2006}.
The finite-time values reported here
refer to the chosen coordinates and
norm; they are not claimed to be
coordinate-independent scalar observables.
Fixed equivalent norms yield the same
asymptotic exponents under the usual
boundedness assumptions, although
finite-time estimates can retain
a norm dependence.

\subsection{Numerical implementation and accuracy assessment}
\label{subsec:numerical_accuracy}

The coupled orbit and tangent equations
are integrated using the adaptive
DOP853 method.
The static seed is first integrated
numerically, and the resulting functions
$K(r)$ and $\Delta(r)$ are represented
by quintic interpolating splines on
a uniform grid of 24001 points spanning
$2.05\leq r/M\leq500$.
Derivatives of these same interpolants
are used to evaluate the Hamiltonian
vector field.

The Jacobian for the tangent equations
is evaluated by centered differences
of that vector field:
\begin{equation}
\begin{split}
(\mathbf{J}_f)_{ij}
&\simeq
\frac{
f_i(y+h_j\hat e_j)-f_i(y-h_j\hat e_j)
}{2h_j},\\
h_j&=\eta D_{jj},
\end{split}
\label{eq:numerical_jacobian}
\end{equation}
where $\hat e_j$ is a coordinate
basis vector.
The baseline integrations use relative
and absolute tolerances $10^{-11}$ and
$10^{-12}$, a maximum step of $1.5M$,
a renormalization interval of $100M$,
and $\eta=10^{-5}$.
These settings control distinct error
sources and should not be interpreted
as a single accuracy parameter.

Integrations are terminated upon reaching
$r/M=2.15$ or $490$, or either polar
boundary $\theta=0.025$ and
$\theta=\pi-0.025$.
These boundaries delimit the computational
domain and are not interpreted as
physical capture or escape criteria.

The primary constraint diagnostic is
\begin{equation}
\delta_H(\tau)
=\left|H(\tau)+\frac{1}{2}\right|.
\label{eq:mass_shell_error}
\end{equation}
For $\beta=0$, we additionally evaluate
\begin{equation}
\delta_Q(\tau)=|Q(\tau)-Q(0)|.
\label{eq:neutral_Q_error}
\end{equation}
For $\beta\neq0$, the corresponding
plotted quantity is instead
\begin{equation}
\Delta Q_{\rm ref}(\tau)
=|Q_{\rm ref}(\tau)-Q_{\rm ref}(0)|,
\label{eq:charged_Q_change}
\end{equation}
which describes dynamical variation
rather than violation of a conserved
constraint.

Since $E$ and $L$ are fixed parameters
of the reduced system, their constancy
is built into the formulation and is
not an independent test of integration
accuracy.

In the comparisons labeled ``refined'',
the integration tolerances are reduced
by a factor of five, while the maximum
step, the tangent-vector renormalization
interval, and $\eta$ are each reduced
by a factor of two.
This joint refinement tests the
sensitivity of the overall result
but does not identify individual
error sources.

Separately varying these settings,
the integration duration, and the
initial tangent direction would provide
a more detailed convergence assessment.
The background interpolation grid is
unchanged in the refined calculations,
so errors shared by those interpolants
are not assessed by refining the
orbit integrator alone.

For additional diagnostics, logarithmic
growth rates over late-time intervals
are evaluated as
\begin{equation}
\lambda_{[T_1,T_2]}
=
\frac{\Lambda(T_2)-\Lambda(T_1)}
{T_2-T_1}.
\label{eq:block_growth_rate}
\end{equation}
In the density comparison, the second
half of each standard integration is
divided into four equal intervals.
The resulting block values characterize
temporal variation along one trajectory;
they are not independent statistical
samples or confidence intervals.

Pointwise agreement between two long
chaotic trajectories is not a suitable
convergence requirement.
The relevant comparisons concern the
persistence of the dynamical
classification, the resolved section
structures, and the stability of
growth-rate estimates.

A trajectory is retained in quantitative
parameter-map comparisons only if it
reaches the requested duration and
satisfies
\begin{equation}
\max_{\tau\in\mathcal{T}_{\rm rec}}
\left|H(\tau)+\frac12\right|
\leq10^{-7},
\label{eq:constraint_quality_gate}
\end{equation}
where $\mathcal{T}_{\rm rec}$ contains
the recorded solver nodes and output
samples.
This is a numerical-quality criterion,
not a threshold for identifying chaos.

The main integrations extend to
$T=8\times10^4M$, while the initial
parameter maps use $T=2\times10^4M$.
The low-growth candidate identified
in Sec.~5 is subsequently integrated
to $T=1.6\times10^5M$ with both
standard and refined settings.

The maps represent finite-time growth
estimates rather than binary
classifications of chaos.
Small mass-shell residuals provide
a consistency check on the orbital
evolution, but do not by themselves
establish convergence of the tangent
dynamics or prove chaotic motion.

\section{Magnetic-field-induced chaos}
\label{sec:magnetic_chaos}

We first examine how the magnetic coupling modifies the orbital dynamics
while keeping the King-type halo background fixed. Unless stated otherwise,
the parameters in this section are
\begin{equation}
\begin{split}
\frac{a}{M}&=0.9,\qquad \frac{r_s}{M}=10,\\
\rho_s M^2&=10^{-6},\qquad
E=0.97,\qquad \frac{L}{M}=3.8.
\end{split}
\label{eq:section4_parameters}
\end{equation}
The representative trajectories start at $r_0=10M$ with
$\theta_0=\pi/2$ and $p_{r0}=0$. For each value of the magnetic
coupling $\beta$, the positive initial polar momentum is determined from
the timelike mass-shell constraint. The comparison therefore follows the
fixed-canonical-constant prescription introduced in Sec.~3.

We consider
\begin{equation}
\beta=0,\quad 0.005,\quad 0.02,\quad 0.04,
\end{equation}
with a common integration duration
\begin{equation}
T=8\times10^4M.
\end{equation}

\subsection{Regular and chaotic structures on the surface of section}
\label{subsec:magnetic_sections}

\begin{figure*}[t]
\centering
\includegraphics[width=0.96\textwidth]{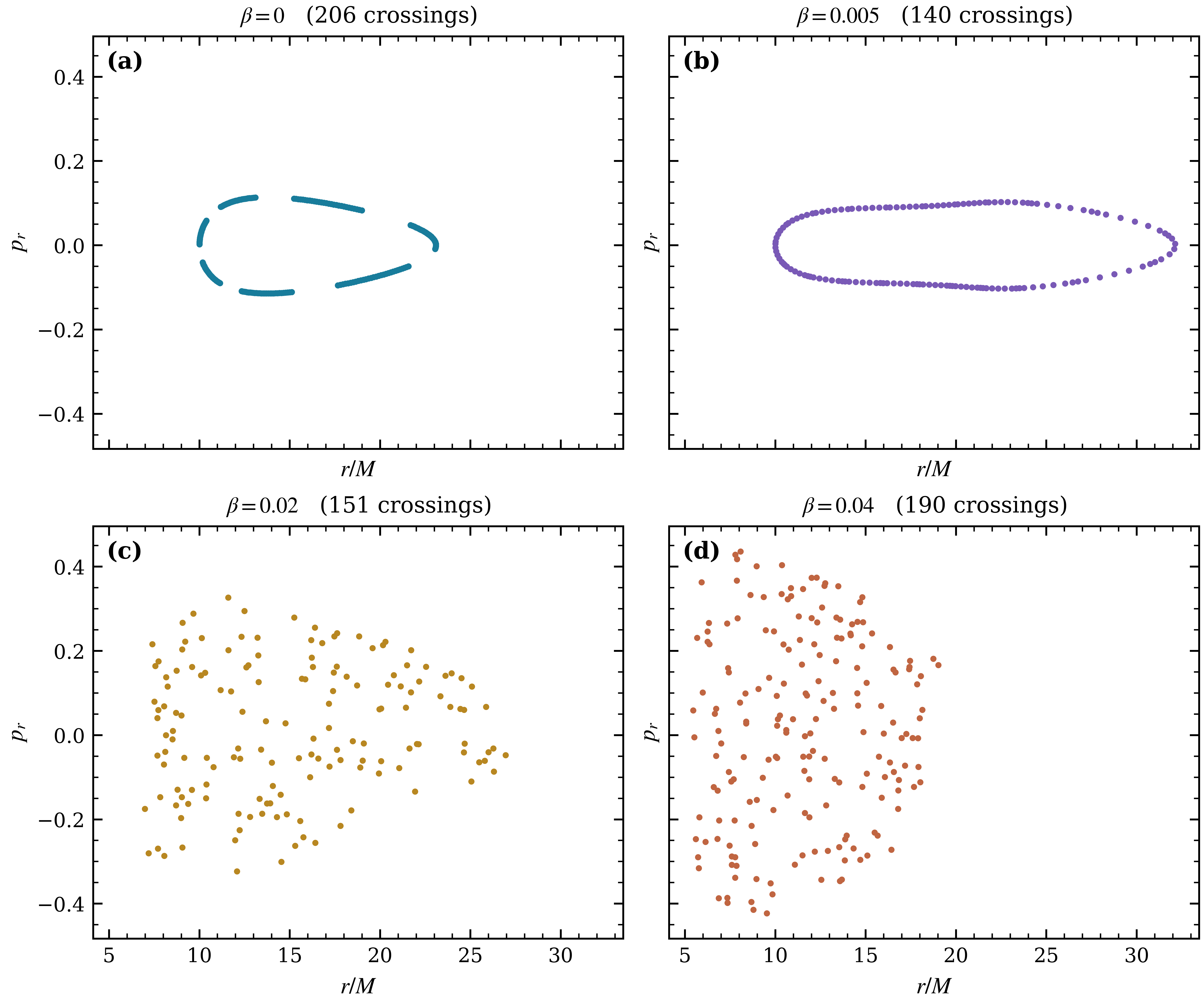}
\caption{
Poincar\'e sections in the $(r,p_r)$ plane for
$\beta=0$, $0.005$, $0.02$, and $0.04$ at fixed King-type halo
parameters. The initial radius is $r_0=10M$, and the remaining parameters
are given in Eq.~\eqref{eq:section4_parameters}. Intersections satisfy
$\theta=\pi/2$ and $\dot{\theta}>0$, with the initial crossing omitted.
All panels use common coordinate limits. The cases $\beta=0$ and $0.005$
show regular closed structures, whereas $\beta=0.02$ and $0.04$ exhibit
scattered two-dimensional sections.
}
\label{fig:magnetic_sections}
\end{figure*}

Figure~\ref{fig:magnetic_sections} shows the corresponding Poincar\'e
sections in the $(r,p_r)$ plane. The four panels are displayed with common
coordinate limits in order to allow a direct comparison of the phase-space
structures.

For $\beta=0$, the intersections remain confined to a smooth
one-dimensional invariant curve, as expected from the Hamilton--Jacobi
separability of the zero-coupling system established in
Sec.~\ref{subsec:integrable_limit}. The finite gaps visible along the curve
are caused by incomplete sampling over the finite integration time and
should not be interpreted as resonance islands or stochastic layers.

For $\beta=0.005$, the section remains a narrow closed curve, although its
shape and radial extent differ from those of the zero-field reference case.
This result demonstrates that breaking the neutral Carter-type integral by
introducing a nonzero magnetic interaction does not automatically imply
chaotic motion. Regular invariant structures can survive in the
magnetically coupled system.

A qualitatively different behavior appears at $\beta=0.02$ and $0.04$.
The intersections no longer follow resolved one-dimensional curves but
instead populate two-dimensional regions of the section. The transition is
especially clear in the $\beta=0.04$ case, for which the orbit spans a
broader range of radial momentum. These scattered structures, when combined
with the tangent-vector growth diagnostics discussed below, support chaotic
motion for the selected trajectories.

The four integrations produce 206, 140, 151, and 190 positive-direction
equatorial crossings for $\beta=0$, $0.005$, $0.02$, and $0.04$,
respectively. The different crossing numbers partly reflect changes in the
orbital timescale and should therefore not be interpreted as a direct
measure of the strength of chaos. Likewise, neither the radial extent nor
the apparent area occupied by a Poincar\'e point cloud alone provides a
quantitative instability diagnostic.

\subsection{Perturbation growth and numerical consistency}
\label{subsec:magnetic_indicators}

\begin{figure*}[t]
\centering
\includegraphics[width=0.98\textwidth]{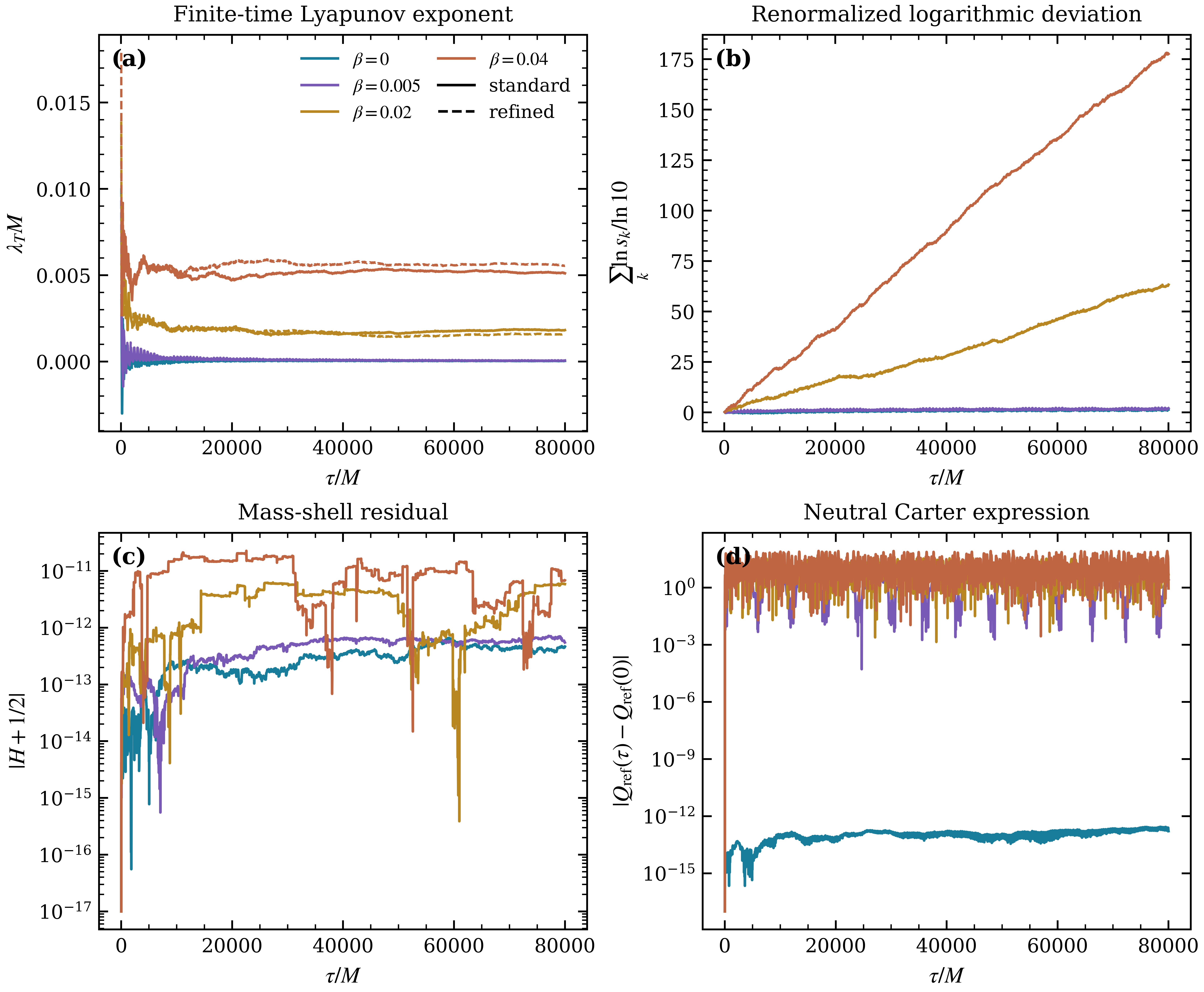}
\caption{
Dynamical and numerical diagnostics for the four trajectories shown in
Fig.~\ref{fig:magnetic_sections}.
(a) Finite-time Lyapunov estimates; solid and dashed curves denote standard
and refined integrations, respectively.
(b) Renormalized logarithmic fast Lyapunov indicator obtained from the
standard integrations.
(c) Absolute mass-shell residual $|H+1/2|$.
(d) Variation
$|Q_{\rm ref}(\tau)-Q_{\rm ref}(0)|$ of the neutral Carter expression.
The latter is a numerical conservation test only for $\beta=0$.
}
\label{fig:magnetic_diagnostics}
\end{figure*}

Figure~\ref{fig:magnetic_diagnostics} compares the finite-time Lyapunov
exponent, the accumulated logarithmic deviation, the mass-shell residual,
and the variation of the neutral Carter expression.

For $\beta=0$ and $0.005$, the finite-time Lyapunov estimates decrease
toward zero after the initial transient. The accumulated logarithmic
deviation also remains small compared with the stronger-coupling cases.
These behaviors are consistent with the regular invariant curves shown in
Fig.~\ref{fig:magnetic_sections}.

For $\beta=0.02$ and $0.04$, the behavior changes qualitatively. The
finite-time Lyapunov estimates remain at sustained positive levels at late
times, and the accumulated logarithmic deviations grow approximately
linearly. The stronger-coupling trajectory exhibits the larger late-time
growth rate. Because the FLI used here and the finite-time Lyapunov estimate
are constructed from the same accumulated tangent-vector growth, they
should not be regarded as independent chaos indicators. Their physical
interpretation is instead strengthened by their agreement with the
Poincar\'e sections.

The dashed curves in Fig.~\ref{fig:magnetic_diagnostics}(a) correspond to
the refined integrations. The refinement preserves the qualitative
separation between the two low-growth trajectories and the two
positive-growth trajectories, and it also preserves the larger growth rate
of the $\beta=0.04$ case relative to $\beta=0.02$. Small quantitative
differences between the standard and refined curves remain, however, so
the individual finite-time exponents should not be interpreted as
high-precision asymptotic Lyapunov exponents.

The mass-shell residuals remain at approximately the
$10^{-12}$--$10^{-11}$ level over most of the integrations, providing a
strong consistency check on the orbital evolution. This constraint
accuracy does not by itself guarantee convergence of the tangent dynamics,
which is why the numerical refinement comparison is performed separately.

The lower-right panel provides an additional useful diagnostic. At
$\beta=0$, the neutral Carter expression is conserved to numerical
precision. At nonzero coupling, its variation becomes substantial.
Importantly, a large variation also occurs for the regular
$\beta=0.005$ trajectory. The nonconservation of the neutral Carter
expression therefore diagnoses the loss of that particular integral but
is not, by itself, a criterion for chaos.

\subsection{Dependence on magnetic coupling and initial radius}
\label{subsec:magnetic_scan}

The representative trajectories above demonstrate a transition from
regular to chaotic behavior as the magnetic coupling is increased, but
they do not establish whether the transition occurs at a universal value
of $\beta$. We therefore perform a finite-time scan over three initial
radii,
\begin{equation}
\frac{r_0}{M}\in\{10,15,20\},
\label{eq:magnetic_scan_radii}
\end{equation}
and nine magnetic couplings,
\begin{equation}
\begin{split}
\beta\in\{&
0,0.0025,0.005,0.0075,0.01,\\
&0.015,0.02,0.03,0.04\}.
\end{split}
\label{eq:magnetic_scan_values}
\end{equation}
Each trajectory is integrated to
\begin{equation}
T=2\times10^4M.
\end{equation}

\begin{figure*}[t]
\centering
\includegraphics[width=0.96\textwidth]{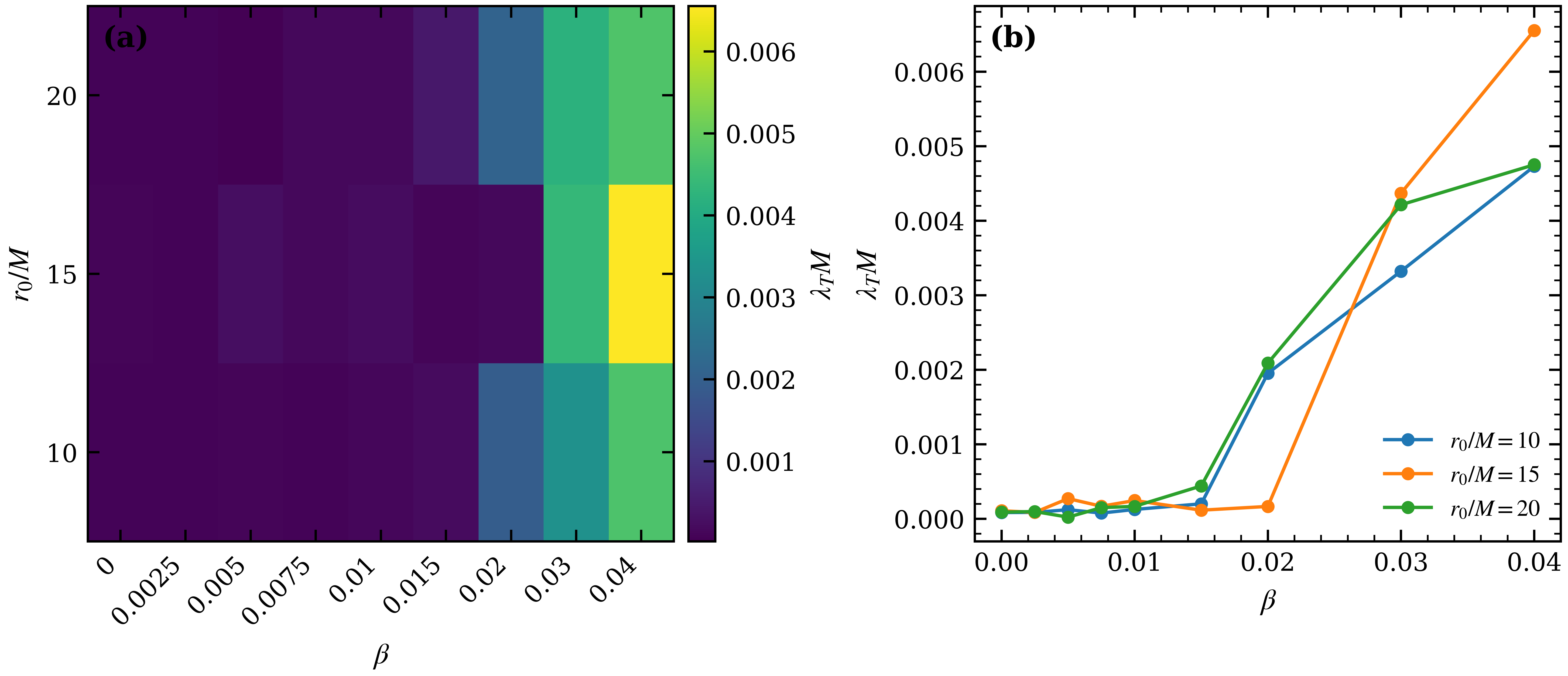}
\caption{
Finite-time magnetic-coupling survey at $T=2\times10^4M$.
(a) Parameter map of $\lambda_T M$ for three initial radii and nine
magnetic couplings. The horizontal coordinate represents the discrete
sampled values of $\beta$.
(b) The same finite-time Lyapunov estimates plotted against the physical
coupling values. Connecting lines guide the eye. The location of the
strong increase in orbital instability depends on the initial radius,
demonstrating that no universal transition coupling is identified by this
scan.
}
\label{fig:magnetic_scan}
\end{figure*}

Figure~\ref{fig:magnetic_scan} shows the resulting finite-time Lyapunov
estimates as both a discrete parameter map and a set of curves plotted
against the physical coupling $\beta$.

At small $\beta$, all three initial-radius families display small
finite-time growth rates. A pronounced increase appears as the coupling
enters the stronger-field regime, but the location of this increase depends
on the initial radius. For $r_0=10M$ and $20M$, the growth rate rises
strongly between the sampled values $\beta=0.015$ and $0.02$. The
$r_0=15M$ family behaves differently: its growth remains small at
$\beta=0.02$ and increases sharply only at larger coupling.

The ordering with initial radius is itself nonmonotonic. In particular,
at $\beta=0.04$ the $r_0=15M$ trajectory has the largest finite-time
exponent among the three sampled radii, while the $r_0=10M$ and $20M$
trajectories have lower and comparable values.

These results show that the regular-to-chaotic transition cannot be
described by a single magnetic-coupling threshold independent of the
initial-condition family. The map should therefore be interpreted as a
finite-time survey rather than as a sharply resolved phase diagram.
Small finite-time exponents may correspond to regular motion, weak chaos,
or sticky trajectories, and longer integrations are required to distinguish
these possibilities in marginal cases.

Taken together, the Poincar\'e sections and tangent-vector diagnostics show
that sufficiently strong magnetic coupling can generate chaotic motion in
the adopted King-halo background, while regular trajectories remain possible
at nonzero coupling. The initial-radius scan further demonstrates that this
transition depends sensitively on the selected orbit family.

\section{Influence of the King-type dark matter halo}
\label{sec:halo_effects}

Having established the magnetic-field-induced transition above, we now
investigate how the King-type halo modifies the charged-particle dynamics
at fixed asymptotic magnetic coupling. We set
\begin{equation}
\begin{split}
\frac{a}{M}&=0.9,\qquad \frac{r_s}{M}=10,\\
E&=0.97,\qquad \frac{L}{M}=3.8,\qquad
\beta=0.02,
\end{split}
\end{equation}
and consider
\begin{equation}
\rho_sM^2
\in
\left\{
0,\,
10^{-6},\,
10^{-5},\,
5\times10^{-5}
\right\}.
\end{equation}

The case $\rho_s=0$ provides the magnetized Kerr reference. For each halo
density, the initial polar momentum is recalculated from the mass-shell
constraint while $E$, $L$, and the coordinate initial conditions are held
fixed. The resulting comparison therefore concerns a prescribed family at
fixed canonical constants rather than identical initial local four-velocities.

\subsection{Density dependence of chaotic trajectories}

\begin{figure*}[t]
\centering
\includegraphics[width=0.96\textwidth]{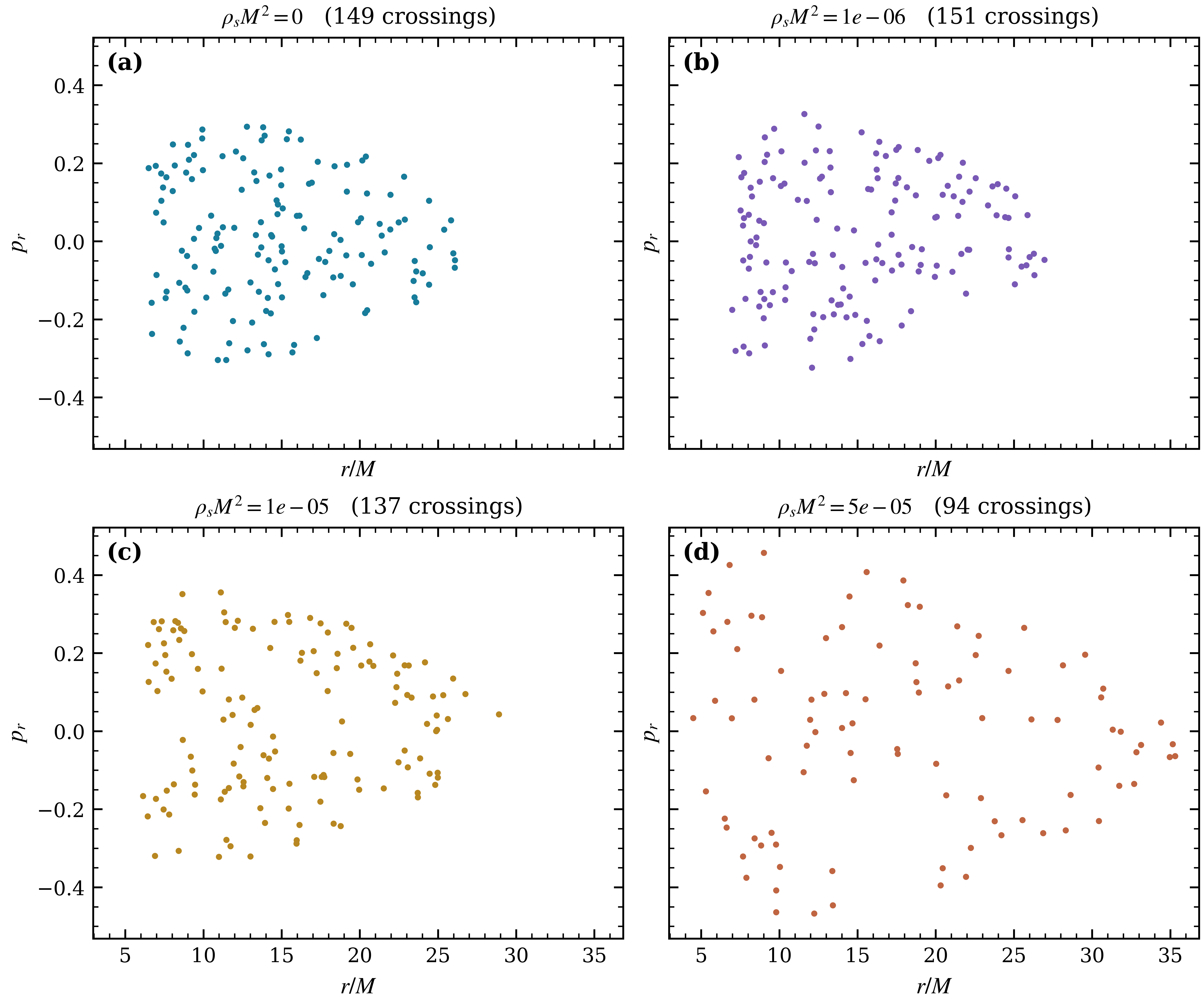}
\caption{
Poincar\'e sections for
$\rho_sM^2=0$, $10^{-6}$, $10^{-5}$, and
$5\times10^{-5}$ at fixed $\beta=0.02$, $r_0=10M$, and
$T=8\times10^4M$. All panels use common coordinate limits.
The magnetized Kerr reference is already chaotic, while increasing halo
density modifies the radial and momentum domains explored by the
trajectory. The corresponding numbers of equatorial crossings are
149, 151, 137, and 94.
}
\label{fig:halo_sections}
\end{figure*}

We first consider $r_0=10M$ and integrate each trajectory to
$T=8\times10^4M$. Figure~\ref{fig:halo_sections} shows the corresponding
Poincar\'e sections.

The magnetized Kerr reference already exhibits a scattered
two-dimensional section, showing that the orbit is chaotic even without a
halo. The nonzero-density cases retain the same broad chaotic morphology,
but the sampled phase-space domain changes as the halo density is increased.
The higher-density trajectories explore a broader range of $p_r$, and the
largest-density case also reaches substantially larger radial excursions.
These changes show that the halo modifies an already existing
magnetically induced chaotic regime rather than generating chaos from an
otherwise regular reference orbit.

The new integrations yield 149, 151, 137, and 94 positive-direction
equatorial crossings for
$\rho_sM^2=0$, $10^{-6}$, $10^{-5}$, and $5\times10^{-5}$,
respectively. Because the integrations all have the same duration, the
different numbers of crossings also reflect density-dependent changes in
the orbital timescale. The number of points appearing in a section should
therefore not be interpreted as a measure of the strength of chaos.

\begin{figure*}[t]
\centering
\includegraphics[width=0.96\textwidth]{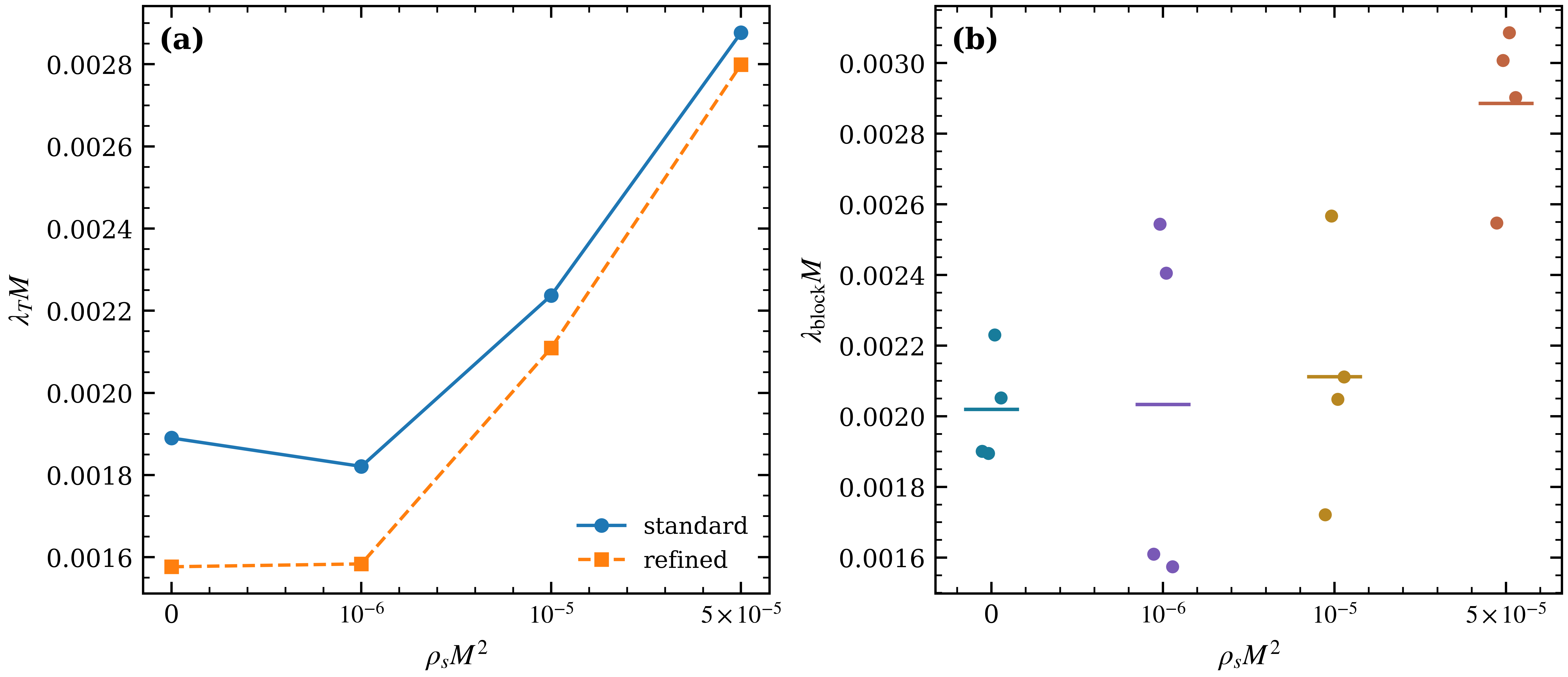}
\caption{
Dependence of the finite-time instability measures on the halo density
for $r_0=10M$ and $\beta=0.02$.
(a) Final finite-time Lyapunov estimates from the standard and refined
integrations at $T=8\times10^4M$.
(b) Growth rates evaluated in four equal intervals covering the second
half of each standard integration. Horizontal segments denote the mean of
the four block values for each density. These block values are correlated
measurements along a single trajectory and are not statistical error bars.
}
\label{fig:halo_density_comparison}
\end{figure*}

The corresponding finite-time Lyapunov estimates are summarized in
Table~\ref{tab:halo_lyapunov} and Fig.~\ref{fig:halo_density_comparison}.

\begin{table}[t]
\centering
\caption{
Finite-time Lyapunov estimates at
$T=8\times10^4M$ for $r_0=10M$ and $\beta=0.02$.
}
\label{tab:halo_lyapunov}
\begin{tabular}{ccc}
\hline
$\rho_sM^2$ &
Standard $\lambda_T M$ &
Refined $\lambda_T M$ \\
\hline
$0$ &
$1.890084\times10^{-3}$ &
$1.576617\times10^{-3}$ \\
$10^{-6}$ &
$1.820945\times10^{-3}$ &
$1.583535\times10^{-3}$ \\
$10^{-5}$ &
$2.237116\times10^{-3}$ &
$2.109251\times10^{-3}$ \\
$5\times10^{-5}$ &
$2.876642\times10^{-3}$ &
$2.799279\times10^{-3}$ \\
\hline
\end{tabular}
\end{table}

The two higher-density cases have larger full-interval growth rates than
the vacuum and weak-halo cases in both numerical settings. Relative to the
vacuum reference, the standard calculation increases by approximately
$18.4\%$ and $52.2\%$ for $\rho_sM^2=10^{-5}$ and
$5\times10^{-5}$, respectively. The corresponding refined increases are
approximately $33.8\%$ and $77.6\%$.

The weak-halo case behaves differently. At $\rho_sM^2=10^{-6}$, the
standard estimate is approximately $3.7\%$ below the vacuum value, whereas
the refined estimate is approximately $0.4\%$ above it. The ordering thus
changes under refinement, and the present calculation does not resolve
whether such a weak halo enhances or suppresses the finite-time instability.

Figure~\ref{fig:halo_density_comparison}(b) shows growth rates evaluated in
four intervals covering the second half of each standard integration. The
spread of these block values demonstrates that the local finite-time growth
varies along a single chaotic trajectory. They are correlated dynamical
measurements rather than independent statistical samples or error bars.

\subsection{Dependence on the initial radius}

\begin{figure}[t]
\centering
\includegraphics[width=\linewidth]{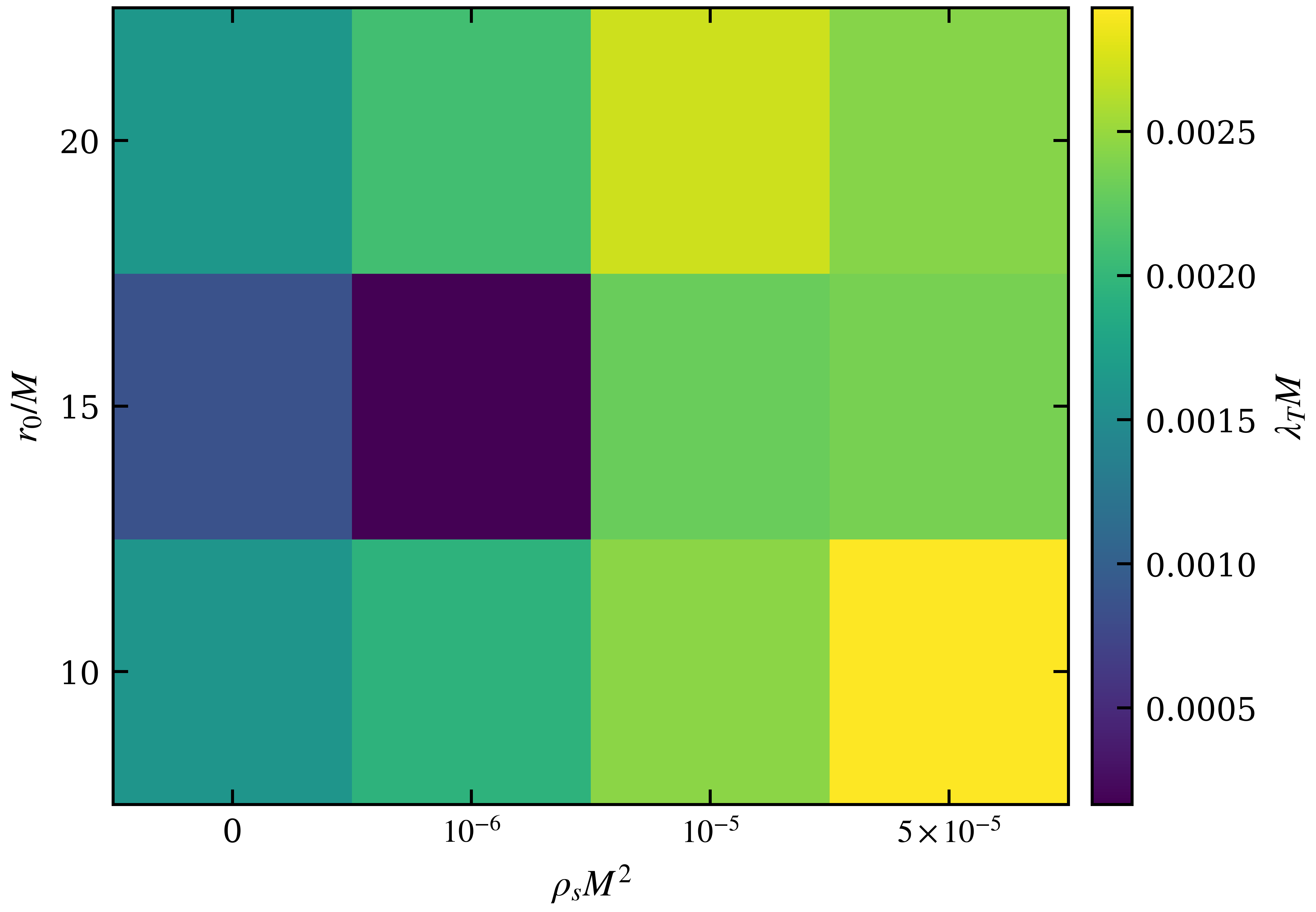}
\caption{
Finite-time Lyapunov map at $\beta=0.02$ and
$T=2\times10^4M$ for three initial radii and four halo densities.
The colors represent the continuous finite-time growth rate $\lambda_T M$,
not a binary regular/chaotic classification. The pronounced low-growth
case at $r_0=15M$ and $\rho_sM^2=10^{-6}$ is examined with a substantially
longer integration below.
}
\label{fig:halo_density_map}
\end{figure}

The density dependence is not universal across orbit families. To examine
this point, we perform a shorter survey with
\begin{equation}
r_0/M\in\{10,15,20\},
\qquad
T=2\times10^4M.
\end{equation}
The resulting finite-time Lyapunov map is shown in
Fig.~\ref{fig:halo_density_map}.

For $r_0=10M$, the finite-time exponent increases strongly toward the two
largest halo densities. The $r_0=20M$ family also exhibits substantially
larger growth rates at higher density, although the dependence is not
strictly monotonic across all sampled points.

The most striking behavior occurs for $r_0=15M$. The weak-halo case
$\rho_sM^2=10^{-6}$ produces a pronounced minimum in the finite-time
growth rate compared with both the vacuum reference and the larger-density
backgrounds. At still higher densities, the exponent increases again.
The response to halo density is therefore strongly dependent on the
location of the initial condition in phase space.

This survey contains only twelve initial states and therefore does not map
the full boundaries of regular and chaotic regions. Its purpose is instead
to demonstrate that the halo response depends on the selected orbit family
and that a simple monotonic relation between $\rho_s$ and chaotic
instability is not supported by the sampled phase space.

\subsection{Extended integration of the low-growth orbit}

To determine whether the low finite-time exponent identified above is only
a short-time transient, we extend the case
\begin{equation}
r_0=15M,\qquad
\rho_sM^2=10^{-6},\qquad
\beta=0.02
\end{equation}
to
\begin{equation}
T=1.6\times10^5M
\end{equation}
using both the standard and refined numerical settings.

As the integration duration is increased, the finite-time Lyapunov estimate
continues to decrease rather than approaching a positive plateau. The
standard and refined final values are also in close agreement. This behavior
is qualitatively different from that of the chaotic trajectories considered
above.

\begin{figure*}[t]
\centering
\includegraphics[width=0.96\textwidth]{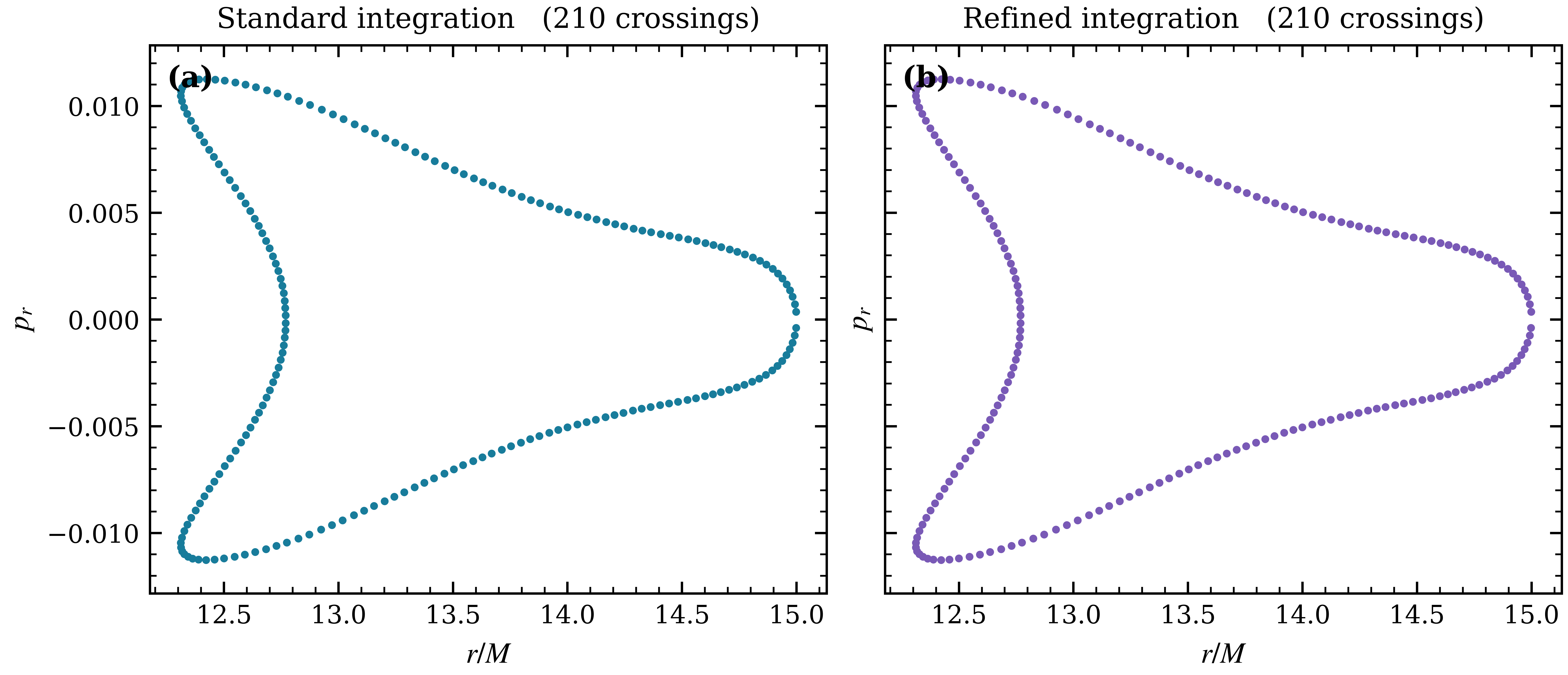}
\caption{
Extended Poincar\'e sections for
$r_0=15M$, $\rho_sM^2=10^{-6}$, and $\beta=0.02$.
The standard and refined integrations both reach
$T=1.6\times10^5M$ and yield 210 equatorial crossings.
The nearly identical closed curves, together with the decreasing
finite-time Lyapunov estimate, support regular quasiperiodic motion over
the computed interval.
}
\label{fig:halo_regular_candidate}
\end{figure*}

Figure~\ref{fig:halo_regular_candidate} shows the corresponding extended
Poincar\'e sections. Both integrations produce 210 positive-direction
equatorial crossings, and the points trace nearly identical smooth closed
curves. The combination of a decreasing finite-time Lyapunov estimate,
a closed Poincar\'e section, and agreement under numerical refinement
provides strong evidence for regular quasiperiodic motion over the computed
interval.

Although these results provide substantially stronger evidence than the
short-time scan alone, a finite integration cannot rigorously exclude an
extremely long sticky chaotic transient. Likewise, one regular trajectory
does not determine the size of the surrounding regular region in phase
space.

\subsection{Numerical checks and physical interpretation}

The numerical results should be interpreted together with the consistency
tests described in Sec.~3. The mass-shell residual remains small throughout
the calculations, while the refined integrations preserve the principal
dynamical classifications. The zero-field controls also recover conservation
of the neutral Carter-type integral to numerical precision.

The magnetized regular trajectory provides an especially useful
counterexample to the interpretation of $Q_{\rm ref}$ as a chaos
diagnostic. Although $Q_{\rm ref}$ varies substantially when the magnetic
coupling is nonzero, the trajectory nevertheless exhibits a closed
Poincar\'e curve and a finite-time Lyapunov estimate decreasing toward zero.
Loss of the neutral Carter integral is therefore not, by itself, sufficient
evidence for chaotic motion.

An additional qualification concerns the interpretation of the density
scan. Within the adopted prescription, changing the halo density modifies
both the effective gravitational geometry and the metric-dependent
electromagnetic potential. Keeping $\beta$ fixed therefore fixes the
asymptotic magnetic coupling but not the locally measured electromagnetic
field at every point in the orbital region. Furthermore, the mass-shell
condition causes the initial polar momentum to change when the halo
background is varied.

The density comparison should consequently be understood as a comparison
within a specified family of effective gravitational and electromagnetic
backgrounds at fixed canonical $E$ and $L$, rather than as an isolated
measurement of a purely gravitational halo correction at identical local
initial conditions.

Within these assumptions, the combined results show that the King-type halo
modulates magnetic-field-induced orbital instability in a strongly
initial-condition-dependent manner. Higher densities enhance the
finite-time instability of the representative chaotic family initialized
at $r_0=10M$, whereas a different orbit at $r_0=15M$ remains regular over
an extended integration. The halo influence therefore cannot be summarized
by a universal monotonic relation between density and chaos strength.

\section{Discussion and conclusions}
\label{sec:conclusions}

We have investigated charged test-particle dynamics
around an effective rotating black hole embedded in
a King-type dark matter halo and exposed to a
prescribed, asymptotically uniform magnetic field.
The analysis combines Hamiltonian evolution,
Poincar\'e sections, finite-time Lyapunov estimates,
and numerical consistency checks to distinguish
regular and chaotic orbital behavior.

An essential feature of the adopted background is
the separability of neutral timelike geodesic motion.
In the zero-coupling limit, the system admits a
Carter-type integral in addition to the energy,
axial angular momentum, and Hamiltonian constraint.
Consequently, the King-type halo does not, by
itself, generate neutral geodesic chaos in this
particular construction. This conclusion concerns
the separable metric adopted here and should not
be generalized to arbitrary rotating halo geometries.

Introducing the electromagnetic interaction changes
the dynamical problem. The neutral Carter expression
is generally no longer conserved, and the numerical
survey identifies trajectories with scattered
Poincar\'e sections and sustained positive
finite-time growth rates. These results support
magnetic-field-induced chaos for the selected
initial conditions. However, a nonzero magnetic
coupling does not imply that every trajectory is
chaotic. Regular orbital structures can persist
within the magnetized system.

The halo-density comparison reveals a clear
dependence on the chosen orbit family. For
$r_0=10M$ and $\beta=0.02$, the two higher-density
backgrounds produce larger full-interval Lyapunov
estimates than the vacuum and weak-halo backgrounds
in both the standard and refined calculations.
At $T=8\times10^4M$, the standard value of
$\lambda_T M$ increases from
$1.890084\times10^{-3}$ in the vacuum case to
$2.876642\times10^{-3}$ at
$\rho_sM^2=5\times10^{-5}$, corresponding to an
increase of approximately $52.2\%$. The refined
value increases from $1.576617\times10^{-3}$ to
$2.799279\times10^{-3}$, corresponding to an
increase of approximately $77.6\%$.
These values characterize the finite-time response
of the selected orbit family and should not be
interpreted as a universal halo correction.

The weak-halo response is less well resolved.
The ordering of the vacuum and
$\rho_sM^2=10^{-6}$ exponents changes under
numerical refinement, so the present calculations
do not establish a definite enhancement or
suppression in this regime.
Likewise, variations among late-time block rates
show that the inferred ordering can depend on
the averaging interval.
These observations limit the precision with which
small density-dependent differences can be
interpreted.

A complementary result is provided by the orbit
with $r_0=15M$, $\rho_sM^2=10^{-6}$,
and $\beta=0.02$.
Extending its integration to $1.6\times10^5M$
produces 210 equatorial crossings tracing a
well-defined closed curve.
The finite-time exponent decreases overall with
increasing duration, and its final value agrees
closely between standard and refined calculations.
Together, these diagnostics support regular,
quasiperiodic motion over the investigated interval.
They do not determine the width of a surrounding
regular island or strictly exclude a much longer
sticky chaotic transient.

The coexistence of these contrasting orbital
responses is the main dynamical outcome of the
study. The halo can increase the instability
of selected chaotic trajectories while other
initial conditions retain regular motion.
The density dependence therefore cannot be
summarized by a single monotonic relation between
halo density and chaos strength.

The calculations also clarify the interpretation
of the diagnostics. The magnetized regular
candidate exhibits appreciable variation of
the neutral Carter expression, demonstrating
that its nonconservation is not sufficient
evidence for chaos. The mass-shell residual
instead measures numerical consistency with
the Hamiltonian constraint.
Furthermore, the finite-time Lyapunov exponent
and the renormalized logarithmic deviation used
here are derived from the same accumulated
tangent-vector growth and should not be counted
as independent confirmations.

Several physical qualifications are necessary.
The rotating geometry is a modified
Newman--Janis-type candidate constructed from
the static halo background. The present work
does not establish that it is an exact rotating
Einstein--matter solution with the intended
dark matter source.
Similarly, the prescribed Wald-type potential
in a nonvacuum background generally requires
supporting currents. The astrophysical origin
and self-consistent distribution of those
currents are not modeled here.
Our conclusions therefore concern test-particle
dynamics within a specified effective
gravitational and electromagnetic background.

The density scan also changes more than the
gravitational potential alone. Because the
electromagnetic potential depends on the metric,
fixing the asymptotic magnetic coupling does not
fix the locally measured field throughout the
orbital region. Moreover, enforcing the
mass-shell constraint changes the initial polar
momentum as the halo density varies.
The comparisons should thus be understood as
responses of a prescribed family at fixed
canonical energy and angular momentum, rather
than as isolated gravitational corrections at
fixed local field and initial four-velocity.

Finally, the calculation adopts conservative
test-particle dynamics and neglects radiation
reaction, dynamical friction, electromagnetic
backreaction, and self-force effects.
It therefore does not directly predict
inspiral waveforms or observable gravitational-wave
signatures. Such applications would require
a dissipative evolution model and a justified
description of the charge and electromagnetic
environment of the orbiting body.

Further work should resolve the neighborhood
of the regular candidate with denser
initial-condition scans and longer integrations.
Separate variations of the integration controls,
tangent-equation differencing, and background
interpolation resolution would help quantify
the remaining numerical sensitivity.
An independent chaos diagnostic and a
self-consistent treatment of the rotating
matter and electromagnetic sources would
provide complementary tests.

Within these stated assumptions, the results
show that a King-type halo modulates
magnetic-field-induced orbital instability
in an initial-condition-dependent manner.
The combination of enhanced growth rates
for selected chaotic trajectories and
persistent regular motion provides a basis
for a more detailed investigation of
phase-space structure in magnetized
black hole environments.

\section*{Funding}

This work is supported by the National Key Research and Development
Program of China (Nos.~2024YFC2207003 and 2025YFE0212600) and related
programs under the China--Brazil Belt and Road Joint Laboratory on
Radio Astronomy Technology.

\section*{Data Availability Statement}
The numerical data generated and analyzed during the current
study are available from the corresponding author on reasonable request.

\section*{Code Availability Statement}
The code used for the numerical calculations in this study is
available from the corresponding author on reasonable request.


\begin{thebibliography}{99}

\bibitem{Kerr1963}
R. P. Kerr,
Gravitational Field of a Spinning Mass as an Example of
Algebraically Special Metrics,
Phys. Rev. Lett. \textbf{11}, 237--238 (1963).
\url{https://doi.org/10.1103/PhysRevLett.11.237}.

\bibitem{Carter1968}
B. Carter,
Hamilton-Jacobi and Schrodinger separable solutions of
Einstein's equations,
Commun. Math. Phys. \textbf{10}, 280--310 (1968).
\url{https://doi.org/10.1007/BF03399503}.

\bibitem{Navarro1997}
J. F. Navarro, C. S. Frenk, and S. D. M. White,
A Universal Density Profile from Hierarchical Clustering,
Astrophys. J. \textbf{490}, 493--508 (1997).
\url{https://doi.org/10.1086/304888}.

\bibitem{King1962}
I. King,
The structure of star clusters. I. An empirical density law,
Astron. J. \textbf{67}, 471 (1962).
\url{https://doi.org/10.1086/108756}.

\bibitem{Gondolo1999}
P. Gondolo and J. Silk,
Dark Matter Annihilation at the Galactic Center,
Phys. Rev. Lett. \textbf{83}, 1719--1722 (1999).
\url{https://doi.org/10.1103/PhysRevLett.83.1719}.

\bibitem{Sadeghian2013}
L. Sadeghian, F. Ferrer, and C. M. Will,
Dark-matter distributions around massive black holes:
A general relativistic analysis,
Phys. Rev. D \textbf{88}, 063522 (2013).
\url{https://doi.org/10.1103/PhysRevD.88.063522}.

\bibitem{Eda2013}
K. Eda, Y. Itoh, S. Kuroyanagi, and J. Silk,
New Probe of Dark-Matter Properties: Gravitational Waves from
an Intermediate-Mass Black Hole Embedded in a Dark-Matter Minispike,
Phys. Rev. Lett. \textbf{110}, 221101 (2013).
\url{https://doi.org/10.1103/PhysRevLett.110.221101}.

\bibitem{Cardoso2022}
V. Cardoso, K. Destounis, F. Duque, R. P. Macedo, and A. Maselli,
Black holes in galaxies: Environmental impact on
gravitational-wave generation and propagation,
Phys. Rev. D \textbf{105}, L061501 (2022).
\url{https://doi.org/10.1103/PhysRevD.105.L061501}.

\bibitem{Xu2018}
Z. Xu, X. Hou, X. Gong, and J. Wang,
Black hole space-time in dark matter halo,
JCAP \textbf{09}, 038 (2018).
\url{https://doi.org/10.1088/1475-7516/2018/09/038}.

\bibitem{Wang2026King}
Y. Wang, M. Liu, and H. Xu,
Quasinormal modes of a Kerr black hole embedded in
a King-type cored dark matter halo,
Eur. Phys. J. C \textbf{86}, 782 (2026).
\url{https://doi.org/10.1140/epjc/s10052-026-16022-2}.

\bibitem{NewmanJanis1965}
E. T. Newman and A. I. Janis,
Note on the Kerr Spinning-Particle Metric,
J. Math. Phys. \textbf{6}, 915--917 (1965).
\url{https://doi.org/10.1063/1.1704350}.

\bibitem{Azreg2014}
M. Azreg-A\"{\i}nou,
Generating rotating regular black hole solutions
without complexification,
Phys. Rev. D \textbf{90}, 064041 (2014).
\url{https://doi.org/10.1103/PhysRevD.90.064041}.

\bibitem{Hansen2013}
D. Hansen and N. Yunes,
Applicability of the Newman-Janis algorithm to black hole
solutions of modified gravity theories,
Phys. Rev. D \textbf{88}, 104020 (2013).
\url{https://doi.org/10.1103/PhysRevD.88.104020}.

\bibitem{Wald1974}
R. M. Wald,
Black hole in a uniform magnetic field,
Phys. Rev. D \textbf{10}, 1680--1685 (1974).
\url{https://doi.org/10.1103/PhysRevD.10.1680}.

\bibitem{Blandford1977}
R. D. Blandford and R. L. Znajek,
Electromagnetic extraction of energy from Kerr black holes,
Mon. Not. R. Astron. Soc. \textbf{179}, 433--456 (1977).
\url{https://doi.org/10.1093/mnras/179.3.433}.

\bibitem{Takahashi2009}
M. Takahashi and H. Koyama,
Chaotic motion of charged particles in an electromagnetic
field surrounding a rotating black hole,
Astrophys. J. \textbf{693}, 472--485 (2009).
\url{https://doi.org/10.1088/0004-637X/693/1/472}.

\bibitem{Kopacek2010}
O. Kop\'{a}\v{c}ek, V. Karas, J. Kov\'{a}\v{r},
and Z. Stuchl\'{\i}k,
Transition from regular to chaotic circulation in
magnetized coronae near compact objects,
Astrophys. J. \textbf{722}, 1240--1259 (2010).
\url{https://doi.org/10.1088/0004-637X/722/2/1240}.

\bibitem{Kopacek2014}
O. Kop\'{a}\v{c}ek and V. Karas,
Inducing chaos by breaking axial symmetry
in a black hole magnetosphere,
Astrophys. J. \textbf{787}, 117 (2014).
\url{https://doi.org/10.1088/0004-637X/787/2/117}.

\bibitem{Stuchlik2016}
Z. Stuchl\'{\i}k and M. Kolo\v{s},
Acceleration of the charged particles due to chaotic
scattering in the combined black hole gravitational field
and asymptotically uniform magnetic field,
Eur. Phys. J. C \textbf{76}, 32 (2016).
\url{https://doi.org/10.1140/epjc/s10052-015-3862-2}.

\bibitem{Sun2021}
W. Sun, Y. Wang, F. Liu, and X. Wu,
Applying explicit symplectic integrator to study chaos
of charged particles around magnetized Kerr black hole,
Eur. Phys. J. C \textbf{81}, 785 (2021).
\url{https://doi.org/10.1140/epjc/s10052-021-09579-7}.

\bibitem{Cao2024}
W. Cao, X. Wu, and J. Lyu,
Electromagnetic field and chaotic charged-particle motion
around hairy black holes in Horndeski gravity,
Eur. Phys. J. C \textbf{84}, 435 (2024).
\url{https://doi.org/10.1140/epjc/s10052-024-12804-8}.

\bibitem{Lu2025}
J. Lu and X. Wu,
Chaos of charged particles near a renormalized group improved
Kerr black hole in an external magnetic field,
Eur. Phys. J. C \textbf{85}, 1122 (2025).
\url{https://doi.org/10.1140/epjc/s10052-025-14853-z}.

\bibitem{Xu2025MOG}
Z. Xu, D. Ma, W. Cao, and K. Li,
Chaotic motion of charged test particles in a Kerr-MOG
black hole with explicit symplectic algorithms,
Eur. Phys. J. C \textbf{85}, 770 (2025).
\url{https://doi.org/10.1140/epjc/s10052-025-14425-1}.

\bibitem{Xu2026Potential}
Z. Xu, D. Ma, and K. Li,
Effects of electromagnetic potential on chaos
in Kerr-MOG black holes,
Eur. Phys. J. C \textbf{86}, 295 (2026).
\url{https://doi.org/10.1140/epjc/s10052-026-15546-x}.

\bibitem{Ahal2025}
Z. Ahal, H. El Moumni, and K. Masmar,
Modeling HF-QPOs in microquasars and AGNs:
charged particles around black holes with CDM halos,
Eur. Phys. J. C \textbf{85}, 1090 (2025).
\url{https://doi.org/10.1140/epjc/s10052-025-14830-6}.

\bibitem{Henon1964}
M. H\'{e}non and C. Heiles,
The applicability of the third integral of motion:
Some numerical experiments,
Astron. J. \textbf{69}, 73 (1964).
\url{https://doi.org/10.1086/109234}.

\bibitem{Benettin1976}
G. Benettin, L. Galgani, and J.-M. Strelcyn,
Kolmogorov entropy and numerical experiments,
Phys. Rev. A \textbf{14}, 2338--2345 (1976).
\url{https://doi.org/10.1103/PhysRevA.14.2338}.

\bibitem{Froeschle1997}
C. Froeschl\'{e}, E. Lega, and R. Gonczi,
Fast Lyapunov indicators. Application to asteroidal motion,
Celest. Mech. Dyn. Astron. \textbf{67}, 41--62 (1997).
\url{https://doi.org/10.1023/A:1008276418601}.

\bibitem{Wu2006}
X. Wu, T.-Y. Huang, and H. Zhang,
Lyapunov indices with two nearby trajectories
in a curved spacetime,
Phys. Rev. D \textbf{74}, 083001 (2006).
\url{https://doi.org/10.1103/PhysRevD.74.083001}.

\bibitem{Skokos2010}
Ch. Skokos,
The Lyapunov Characteristic Exponents and Their Computation,
Lect. Notes Phys. \textbf{790}, 63--135 (2010).
\url{https://doi.org/10.1007/978-3-642-04458-8_2}.
\end{thebibliography}
\end{document}